\documentstyle[12pt,graphicx]{article}                
\def\Journal#1#2#3#4{{#1} {\bf #2}, #3 (#4)}

\def\NPB{{\em Nucl. Phys.} B}
\def\PLB{{\em Phys. Lett.}  B}
\def\PRL{\em Phys. Rev. Lett.}
\def\PRD{{\em Phys. Rev.} D}

\begin{document}
%This is dvips(k) 5.86 Cobegin{document}
\baselineskip 18pt
%t
\def\today{\ifcase\month\or
 January\or February\or March\or April\or May\or June\or
 July\or August\or September\or October\or November\or December\fi
 \space\number\day, \number\year}
\def\thebibliography#1{\section*{References\markboth
 {References}{References}}\list
 {[\arabic{enumi}]}{\settowidth\labelwidth{[#1]}
 \leftmargin\labelwidth
 \advance\leftmargin\labelsep
 \usecounter{enumi}}
 \def\newblock{\hskip .11em plus .33em minus .07em}
 \sloppy
 \sfcode`\.=1000\relax}
\let\endthebibliography=\endlist
\def\lsim{\ ^<\llap{$_\sim$}\ }
\def\gsim{\ ^>\llap{$_\sim$}\ }
\def\r2{\sqrt 2}
\def\beq{\begin{equation}}
\def\eeq{\end{equation}}
\def\beqn{\begin{eqnarray}}
\def\eeqn{\end{eqnarray}}
\def\rmuu{\gamma^{\mu}}
\def\rmud{\gamma_{\mu}}
\def\PL{{1-\gamma_5\over 2}}
\def\PR{{1+\gamma_5\over 2}}
\def\sinW2{\sin^2\theta_W}
\def\AEM{\alpha_{EM}}
\def\mul{M_{\tilde{u} L}^2}
\def\mur{M_{\tilde{u} R}^2}
\def\mdl{M_{\tilde{d} L}^2}
\def\mdr{M_{\tilde{d} R}^2}
\def\mz2{M_{z}^2}
\def\c2b{\cos 2\beta}
\def\au{A_u}         
\def\ad{A_d}
\def\cob{\cot \beta}
\def\v#1{v_#1}
\def\tb{\tan\beta}
\def\epem{$e^+e^-$}
\def\KK{$K^0$-$\bar{K^0}$}
\def\wi{\omega_i}
\def\xj{\chi_j}
\def\Wmu{W_\mu}
\def\Wnu{W_\nu}
\def\m#1{{\tilde m}_#1}
\def\mH{m_H}
\def\mw#1{{\tilde m}_{\omega #1}}
\def\mx#1{{\tilde m}_{\chi^{0}_#1}}
\def\mc#1{{\tilde m}_{\chi^{+}_#1}}
\def\mwi{{\tilde m}_{\omega i}}
\def\mxi{{\tilde m}_{\chi^{0}_i}}
\def\mci{{\tilde m}_{\chi^{+}_i}}
\def\mz{M_z}
\def\sw{\sin\theta_W}
\def\cw{\cos\theta_W}
\def\cb{\cos\beta}
\def\sb{\sin\beta}
\def\rwi{r_{\omega i}}
\def\rxj{r_{\chi j}}
\def\rfp{r_f'}
\def\Kik{K_{ik}}
\def\Fq2{F_{2}(q^2)}
\def\f{\({\cal F}\)}
\def\d1{{\f(\tilde c;\tilde s;\tilde W)+ \f(\tilde c;\tilde \mu;\tilde W)}}
%%%%%%%%%%%%%%%%%%%%%%%%%%%%%%%%%%
\def\tw{\tan\theta_W}
\def\sec2w{sec^2\theta_W}
%%%%%%%%%%%%%%%%%%%%%%%%%%%%%%%%%%

\begin{titlepage}

\begin{center}
{\large {\bf Decays of Higgs to $b\bar b$, $\tau \bar \tau$ and
$c\bar c$ 
as Signatures of Supersymmetry and CP Phases}}\\
% Branching Ratio $R_{b/\tau}$ of the Lightest
%Higgs with CP Phases}}\\
\vskip 0.5 true cm
\vspace{2cm}
\renewcommand{\thefootnote}
{\fnsymbol{footnote}}
 Tarek Ibrahim$^{a,b}$ and Pran Nath$^{b}$  
\vskip 0.5 true cm
\end{center}

\noindent
{a. Department of  Physics, Faculty of Science,
University of Alexandria,}\\
{ Alexandria, Egypt\footnote{: Permanent address of T.I.}}\\ 
{b. Department of Physics, Northeastern University,
Boston, MA 02115-5000, USA } \\
%\footnote{ $\dagger$ : Permanent address}
\vskip 1.0 true cm
\centerline{\bf Abstract}
\medskip
The branching ratio of the lightest Higgs decay
into $b\bar b$, $\tau\bar  \tau$  and $c\bar c$ is sensitive to supersymmetric
effects. We include in this work the effects of CP phases on the
Higgs decays. Specifically we compute the deviation of the 
CP phase dependent branching ratio from the standard model
result. The analysis includes the full one loop corrections
of fermion masses  including
CP phases involving the gluino, the chargino and the neutralino exchanges.
 The analysis shows that the supersymmetric effects with CP phases
 can change the branching ratios by as much as 100\% 
 for the lightest Higgs decay into $b\bar b$ and $\tau\bar  \tau$ 
 with similar results holding for the heavier Higgs decays.
 A detailed analysis is also given of the effects of CP phases on
 the Higgs  decays into $c\bar c$. The deviations of $R_{b/\tau}$ and 
 $R_{b/c}$ from the standard model result are investigated as a 
 possible signature of supersymmetry and CP effects. Thus a measurement
 of the decays of the  higgs into $b\bar b$, $\tau\bar \tau$  and $c\bar c$ may 
provide important clues regarding the existence of supersymmetry and
CP phases.
\end{titlepage}
\section{Introduction}
Is is known that supersymmetric contributions can significantly 
affect the Higgs decays into $b\bar b$,  
$\tau\bar \tau$ and $c\bar c$\cite{babu1998}. We compute here the effects 
of CP phases on these arising from soft breaking parameters.
 Thus, for example, the simplest supergravity unified model
 mSUGRA\cite{msugra}, whose soft breaking sector is defined by the parameters
 $m_0, m_{\frac{1}{2}}, A_0, \tan\beta$ (where $m_0$ is the universal scalar mass,
 $m_{\frac{1}{2}}$ is the universal gaugino mass, $A_0$ is the universal
 trilinear coupling and $\tan\beta =<H_2>/<H_1>$ where $H_2$ gives
 mass to the up quark and $H_1$ gives mass  to the down quark and
 the lepton) can accommodate two CP violating phases which can be 
 chosen to be the phase of the Higgs mixing parameter 
 $\mu$ and the phase of $A_0$. 
Extended versions of mSUGRA including nonuniversalities can accommodate
more phases. In this analysis we will consider 
the more general case of the supersymmetric standard model (MSSM)
which allows for several CP phases. Thus, for example, we will
allow the gaugino masses to be in general nonuniversal so that 
$\tilde m_i=|\tilde m_i|e^{i\xi_i} (i=1,2,3)$.
 In the analysis we use the
results recently obtained regarding the effects of CP phases
on the third generation quark and lepton masses\cite{in2003}. 
The analysis of Ref.\cite{in2003} extends the analyses of 
Refs.~\cite{Hall:1993gn,carena2000,carena2003} where large corrections
to third generation masses and specifically to the b quark 
mass, but without inclusion of CP phases, were found.
The corrections to the quark and lepton masses are of considerable 
importance in the analysis of Yukawa unification\cite{arason,shafi,baer}. 
In this paper we utilize these corrections to study their effects on
the Higgs decays into $b\bar b$ and $\tau \bar \tau$.

The analyses which include CP phases must be constrained by the 
experimental upper limits on
the electric dipole moment for the electron and for the neutron
which are very stringent.  Thus for the electron the current
experimental limit for the magnitude of electron edm is 
$d_e<4.3\times 10^{-27}ecm$\cite{eedm}
while for the neutron it is $d_n<6.5\times 10^{-26}ecm$\cite{nedm}.
There is a similar stringent limit on the edm of the $H_g^{199}$ atom,
i.e., $d_{H_g}<9\times 10^{-28}ecm$\cite{atomic}.  
A variety of techniques have
been discussed in the literature to achieve consistency with
experiment\cite{ellis,na,bdm2,incancel,inmssm,inbrane,chang,olive,inhg199}.
Specifically one finds that it is possible to accommodate large
CP phases and still have consistency with the edm constraints
either via the cancellation mechanism\cite{incancel,inbrane,olive,inhg199} or 
via large phases in the third generation sector\cite{chang}. 
We note in passing the analyses of Refs.\cite{olive,inhg199} also include
$H_g^{199}$ edm constraint and show that this constraint along with the
electron and the neutron edm constraint can be satisfied in the presence
of large phases. Of course, if the phases are large they will affect
a variety of low energy phenomena.  Such effects
have been investigated on a variety of processes.
These include effects on the Higgs sector\cite{pilaftsis,inhiggs},
on $g_{\mu}-2$\cite{ing2}, 
on collider physics\cite{kane,zerwas}, in 
B physics and in flavor violation\cite{masiero1,inhg199,huang,demir2003} and
 a variety of other low energy phenomena. The number of phenomena investigated
 is rather larger and a more complete list can be found in 
 Ref.\cite{insusy02}. 

The outline of the rest of the paper is as follows: In Sec.2 we give
a brief description of the basic formalism needed  for the 
evaluation of the full one loop effects on the decays of the Higgs into 
quarks and leptons including the effects of CP phases.
In Sec.3 we give the effective low energy interaction of the b quark with
the lightest Higgs including loop corrections. The branching ratios
$BR(H_2\rightarrow b\bar b)$ and $BR(H_2\rightarrow \tau\bar \tau)$
are also computed.
In Sec.4 we take the limit of vanishing phases and compare our results
to previous analyses. 
 In Sec.5 we give a numerical analysis of the deviations of the
 branching ratios from the standard model predictions because
 of supersymmetric effects 
 and discuss the sensitivity of these deviations to CP phases.
  Conclusions are given in Sec.6.  In Appendix A we 
  extend the results of Sec.4 to include $H_2\rightarrow c\bar c$
  decay and compute the supersymmetry and CP effects on 
  $\Delta R_{b/c}$.  
Extension of the results to decays of the Higgs bosons $H_1$ and
$H_3$ is given in Appendix B. 
\section{The basic formalism}
At the tree level the b quark couples to the neutral component
of $H_1$ Higgs boson while the coupling to the $H_2$ higgs bosons 
is absent where
\beqn
(H_1)= \left(\matrix{H_1^0\cr
 H{'}_{1}^-}\right),~~
(H_2)= \left(\matrix{H{'}_{2}^+\cr
             H_2^0}\right)
\label{2a}
\eeqn
Loop corrections produce a shift in the $H_1^0$ couplings and generate
a non vanishing effective coupling with $H_2^0$. Thus the effective
Lagrangian would be written as 
\beqn
-{\cal {L}}_{eff}= (h_b+\delta h_b)\bar b_R b_L H_1^0 
+ \Delta h_b \bar b_R b_L H_2^{0*} + H.c.
\label{2b}
\eeqn
where the star on $H_2^{0*}$ is necessary in order to have a 
gauge invariant ${\cal {L}}_{eff}$. The same analysis holds for the
tau-lepton sector where $h_{\tau}$, $\delta h_{\tau}$ and 
$\Delta h_{\tau}$ are used in the Lagrangian involving 
$\bar\tau_R \tau_L$.

The quantities $\delta h_f$ and $\Delta h_f$ receive SUSY QCD and
SUSY electroweak contributions. They are calculated in 
Ref.\cite{in2003} on
mass corrections to lepton and quark masses. In the analysis 
carried out there one finds that the couplings are generally complex
due to CP phases in the soft breaking terms. Electroweak 
symmetry is broken spontaneously by giving vacuum expectation value
to $H_1^0$ and $H_2^0$. Thus one finds
\beqn
(H_1)
 =\frac{1}{\sqrt 2} 
\left(\matrix{v_1+\phi_1+i\psi_1\cr
             H_1^-}\right),~~~
(H_2) 
=\frac{e^{i\theta_H}}{\sqrt 2} \left(\matrix{H_2^+ \cr
             v_2+\phi_2+i\psi_2}\right)
\label{2c}
\eeqn
Inserting in $H_1^0$ and $H_2^0$ one finds 
\beqn
-{\cal {L}}_m = M_b \bar b_R b_L + H.c.
\label{2d}
\eeqn
where 
\beqn
M_b= \frac{h_b v_1}{\sqrt 2} \{ 1+ \frac{\delta h_b}{h_b}
+\frac{\Delta h_b}{h_b} \tan\beta \}
\label{2e}
\eeqn
Here $M_b$ is complex because $\delta h_b$ and $\Delta h_b$ are both
complex. We carry out a redefinition of the b quark field
\beqn
b=e^{i\frac{1}{2}\gamma_5 \chi_b} b',~~~\tan\chi_b 
=\frac{Im M_b}{Re M_b} 
\eeqn
After the redefintion of Eq.(6) the mass term reads  
\beqn
-{\cal {L}}_m = m_b \bar b_R' b_L' + H.c.
\eeqn
where $m_b$ is real and positive and $b'$ is the physical field and
$m_b$ and $h_b$ are related by 
\beqn
h_b = \frac{\sqrt 2 m_b}{v_1} 
[(1+ Re\frac{\delta h_b}{h_b} +Re\frac{\Delta h_b}{h_b} \tan\beta )^2 
+ (Im\frac{ \delta h_b}{h_b} 
+ Im\frac{\Delta h_b}{h_b} \tan\beta )^2]^{-\frac{1}{2}}
\eeqn
 The above  can be approximated by 
 \beqn
 h_b \simeq \frac{\sqrt 2 m_b}{v_1} \frac{1}{(1+ \Delta_b)},~~
\Delta_b =Re\frac{\delta h_b}{h_b} + Re\frac{\Delta h_b}{h_b} \tan\beta
\eeqn
From now on we will drop the prime on b and we will assume that we 
are already in the basis where the b field is the physical field for
the b quark. The other terms in  Eq.~(\ref{2b}) which have
 $\phi_1, \psi_1,\phi_2$ and $\psi_2$  dependence will produce the interaction
 between the b-quark and the mass eigen states of the Higgs fields $H_i$.
 Next we introduce the basis  
 $\phi_1, \phi_2,\psi_{1D},\psi_{2D}$ where 
\beqn
\psi_{1D} = \sin\beta \psi_1 + \cos\beta \psi_2,~~
\psi_{2D} = -\cos\beta \psi_1 + \sin\beta \psi_2
\eeqn
In this basis the field $\psi_{2D}$ is the would be Goldstone field.
By considering one loop contributions  to the Higgs boson masses
and mixings from top-stop, bottom-sbottom, chargino and 
neutralino exchanges with CP violating phases, the mass eigen states
of the Higgs fields would be mixed states of CP even  and CP odd states.
Thus the mass eigen states $H_i$ (i=1,2,3) are related to the 
$\phi_1, \phi_2,\psi_{1D}$ as 
follows\cite{pilaftsis,inhiggs,carena2001,carena2002}
\beqn
\left(\matrix{H_1\cr
             H_2\cr H_3}\right) = R
   \left(\matrix{\phi_1\cr
             \phi_2\cr\psi_{1D}}\right) 
\eeqn
 In the absence of CP phases there is no mixing between
the CP even and the CP odd Higgs and the Higgs mass$^2$ matrix
consists of a $2 \times 2$ matrix for the CP even fields and a 
one element for the CP odd Higgs.
Solving for the eigen values in this case one may choose
the first eigen value to be the heavier mass and the second
eigen value to be the lighter mass. 
In previous analyses where  we considered the effects  of CP phases
on the Higgs sector\cite{inhiggs}, we used the convention that 
in the limit of vanishing
CP phases one has $H_1\rightarrow H, H_2\rightarrow h,
H_3\rightarrow A$\cite{inhiggs}. In the present analysis we
 continue to use the same convention. Thus the lightest Higgs boson 
field  corresponds to $H_2$ in our notation. 
\section{The Decays {\boldmath{
 $H_2\rightarrow b\bar b$ and $H_2\rightarrow \tau\bar \tau$ 
 including effects of CP phases}} }
In this section we study the decay of $H_2$ Higgs into
$b\bar b$ and $\tau\bar \tau$. Consider the b quark first. 
The effective interaction of the b quark with the Higgs mass
eigen states $H_2$ is given by 
\beqn
-{\cal {L}}_{int}^b= \bar b [C_b^S + i\gamma_5 C_b^P]b H_2\nonumber\\
C_b^S= \tilde C_b^S \cos\chi_b -\tilde C_b^P \sin\chi_b\nonumber\\
 C_b^P= \tilde C_b^S \sin\chi_b +\tilde C_b^P \cos\chi_b\nonumber\\
\sqrt 2 \tilde C_b^S = Re (h_b +\delta h_b) R_{21} + [-Im (h_b + \delta h_b) 
 \sin\beta\nonumber\\  + Im (\Delta h_b)\cos\beta ] R_{23}
+ Re (\Delta h_b) R_{22}\nonumber\\
\sqrt 2 \tilde C_b^P = - Im (h_b +\delta h_b) R_{21} + [-Re (h_b + \delta h_b)
\sin\beta\nonumber\\ + Re (\Delta h_b) \cos\beta ]R_{23}
- Im (\Delta h_b) R_{22} 
\label{3a}
\eeqn
In deriving ${\cal {L}}^{b}_{int}$ above, we redefined the quark
field $b$ such that $h_b$ is real and is given by Eq.(8) while the
quark mass $M_b$ is complex before field redefinition.
In  Ref.~\cite{carena2001} the authors 
redefined the fields such that $h_b$ is complex and is given by 
their Eq.(4.5) while the quark mass is real and positive. 
To compare now  our Eq.(12) with Eq.(4.10) of Ref.~\cite{carena2001}
and Eq. (77) of  Ref.~\cite{carena2003}, we have the phase $\chi_b$ of 
the quark mass apparent while $h_b$ in $C^{S}_{b}$ and $C^{P}_{b}$ is real.
In  Eqs.(4.11, 4.12)
of Ref.~\cite{carena2001} and 
 Eq.(78, 79) of Ref.~\cite{carena2003}
  the phase $\chi_b$ is absent because of their choice of phases which
  makes $h_b +\delta h_b +\Delta h_b\tan \beta$ a real and positive 
  quantity and our Eq.(6) would lead
then to a vanishing $\chi_b$. However $h_b$ in these expressions
is complex and this will compensate for the absence of $\chi_b$.
So both methods, the one given here and that of Refs.\cite{carena2001} 
and \cite{carena2003}, lead to the same phenomenolgical results.\\

\begin{figure}
\hspace*{-0.6in}
\centering
\includegraphics[width=9cm,height=4cm]{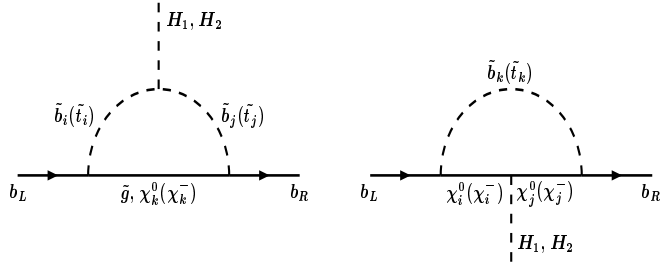}
\caption{One loop contribution to the bottom quark mass involving 
exchange of gluino, charginos and neutralinos in the loop.}
\label{bloop}
\end{figure}
The $\tau$ lepton has similar interactions. Thus 
\beqn  
-{\cal {L}}_{int}^{\tau}= \bar \tau [C_{\tau}^S + i\gamma_5 C_{\tau}^P]b H_2 
\label{3b}
\eeqn
%\label{3.2}
\noindent
where $C_{\tau}^S$ and $C_{\tau}^P$ are given by Eq.~(\ref{3a}) with the
transposition $b\rightarrow \tau$. We are interested in the ratio of the
branching ratios for $H_2\rightarrow \bar b b$ and 
$H_2\rightarrow \bar \tau \tau$, i.e.,  
\beqn
R_{b/\tau} =\frac{BR(H_2\rightarrow \bar b b)}{BR(H_2\rightarrow \bar\tau\tau)}
\label{3c}
\eeqn
Using the interactions given in Eqs.~(\ref{3a})  and ~(\ref{3b}) we can estimate this
to be
\beqn
R_{b/\tau} = 3\frac{m_b^2}{m_{\tau}^2} \frac{S_{\tau}}{S_b} 
\frac{T_{b}}{T_{\tau}} (1+\omega) 
\label{3d}
\eeqn
Here 
\beqn
T_f= (\frac{M_{H_2}^2-4m_f^2}{M_{H_2}^2})^{\frac{3}{2}}
(A_f^S)^2 +
(\frac{M_{H_2}^2-4m_f^2}{M_{H_2}^2})^{\frac{1}{2}}(A_f^P)^2
\eeqn
where
\beqn
A_f^S=\tilde A_f^S \cos\chi_f - \tilde A_f^P \sin\chi_f,~~
A_f^P=\tilde A_f^S \sin\chi_f +\tilde A_f^P \cos\chi_f\nonumber\\
\tilde A_f^S=[(1+Re\frac{\delta h_f}{h_f}) R_{21}  - Im\frac{\delta h_f}{h_f} R_{23} 
\sin\beta + Re\frac{\Delta h_f}{h_f}R_{22}
+ Im\frac{\Delta h_f}{h_f}R_{23}
\cos\beta]\nonumber\\
\tilde A_f^P=
[-Im\frac{\delta h_f}{h_f} R_{21}  - (1+ Re\frac{\delta h_f}{h_f}) R_{23} 
\sin\beta
- Im\frac{\Delta h_f}{h_f}R_{22} + Re\frac{\Delta h_f}{h_f}R_{23}
\cos\beta]
\label{3e}
\eeqn
\beqn
S_f= 
[(1+ Re\frac{\delta h_f}{h_f}  + Re\frac{\Delta h_f}{h_f} \tan\beta)^2
+ (Im\frac{\delta h_f}{h_f}  + Im\frac{\Delta h_f}{h_f} \tan\beta)^2]
\label{3f}
\eeqn
and $(1+\omega)$ is the QCD enhancement factor and is 
given by\cite{gorishnii}
\beqn
(1+\omega) = 1+ 5.67 \frac{\alpha_s}{\pi} + 29.14 \frac{\alpha_s^2}{\pi^2}
\label{3g}
\eeqn
so that $(1+\omega)\simeq 1.25$ for $\alpha_S\simeq 0.12$. 
CP phases enter in Eq.~(\ref{3d}) through $T_f$ and $S_f$ since these 
depend on the phases through the couplings $\Delta h_f$, $\delta h_f$ 
and through the matrix element $R_{2i}$.  Now the CP dependence
of $\Delta h_b$ and  $\delta h_b$  is significantly different from
the CP dependence of $\Delta h_{\tau}$ and $\delta h_{\tau}$ on
phases since, for example, corrections to the b quark mass involve gluino
contributions and are thus sensitive to the phase $\xi_3$ while
the corrections to the $\tau$ lepton mass do not depend on this phase.
Thus one can expect a sensitive dependence of the ratio $R_{b/\tau}$
on the phases. We may compare the above result to the result from
the Standard Model. Here one has 
\beqn
(R_{b/\tau})_{SM} =
 3(\frac{m_b^2}{m_{\tau}^2}  )  
(\frac{m_h^2-4m_b^2}{m_h^2-4m_{\tau}^2})^{\frac{3}{2}}(1+\omega)
\label{3h}
\eeqn
By identifying $m_h$ with $m_{H_2}$ we define the shift due to
supersymmetric effects including the effects due to CP phases as
follows
\beqn
\Delta R_{b/\tau} = \frac{R_{b/\tau} - (R_{b/\tau})_{SM}}
{R_{b/\tau}}
\label{3i}
\eeqn 
As pointed out in Ref.\cite{babu1998} this ratio can be used to 
distinguish the SM Higgs from the lightest SUSY Higgs. The same holds here
except that we also take into account the effects of CP phases.
We will discuss the effects of the CP phases on $\Delta R_{b/\tau}$
numerically in Sec.5. 
It is also of interest to analyze the ratio of the branching ratios
 $H_2\rightarrow b\bar b$ and $H_2\rightarrow c\bar c$. Thus define 
\beqn
R_{b/c} =\frac{BR(H_2\rightarrow \bar b b)}{BR(H_2\rightarrow c\bar c)}
\label{3j}
\eeqn
By repeating the analysis as for the previous case we 
get 
\beqn
R_{b/c} = \frac{m_b^2}{m_c^2} \tan^2 \beta \frac{S_c'}{S_b} 
\frac{T_{b}}{T_c'}  
\label{3k}
\eeqn 
where 
\beqn
T_c'= (\frac{M_{H_2}^2-4m_f^2}{M_{H_2}^2})^{\frac{3}{2}}
(A_c^S)^2 +
(\frac{M_{H_2}^2-4m_f^2}{M_{H_2}^2})^{\frac{1}{2}}(A_c^P)^2
\eeqn
where
\beqn
A_c^S=\tilde A_c^S \cos\chi_c - \tilde A_c^P \sin\chi_c,~~
A_c^P=\tilde A_c^S \sin\chi_c +\tilde A_c^P \cos\chi_c\nonumber\\
 \tilde A_c^S= [(1+Re\frac{\delta h_c}{h_c}) R_{22}  -Im\frac{\delta h_c}{h_c} R_{23} 
\cos\beta
+ Re\frac{\Delta h_c}{h_c}R_{21}
+ Im\frac{\Delta h_c}{h_c}R_{23}
\sin\beta]\nonumber\\ 
\tilde A_c^P=
[-Im\frac{\delta h_c}{h_c} R_{22}  - (1+ Re\frac{\delta h_c}{h_c}) R_{23} 
\cos\beta 
- Im\frac{\Delta h_c}{h_c}R_{21} + Re\frac{\Delta h_c}{h_c}R_{23}
\sin\beta]
\label{3l}
\eeqn
\beqn
S_c'= 
[(1+Re\frac{\delta h_c}{h_c}  + Re\frac{\Delta h_c}{h_c} \cot\beta)^2
+ (Im\frac{\delta h_c}{h_c}  + Im\frac{\Delta h_c}{h_c} \cot\beta)^2]
\label{3m}
\eeqn
We note that the prime is used on $T_c'$ and $S_c'$ for the c quark
case  since it has different couplings. The $\Delta h_c$ and
$\delta h_c$ can be deduced similar to the analysis of $\Delta h_t$
and $\delta h_t$ given in Ref.\cite{in2003}. Explicit results are
given in Appendix A. An analysis similar to the above but for the decay of
the heavier Higgs bosons $H_1$ and $H_3$ are given in Appendix B. 
\section{Limit of Vanishing Phases}
We compare our results now with previous analysis of 
 Ref.\cite{babu1998} by taking the
limit of vanishing phases. In this limit we set 
\beqn
\chi_{f}=0,~~Im(\delta h_f) =0= Im(\Delta h_f)\nonumber\\
R_{11}= \cos\alpha, ~~R_{12}=\sin\alpha\nonumber\\
R_{21} = -\sin\alpha, ~~ R_{22}=\cos\alpha\nonumber\\
R_{13}=R_{23}=0
\label{4a}
\eeqn
Neglecting the masses of b, c and $\tau$ relative to the masses of the
Higgs bosons,  one obtains from Eqs.~(\ref{3d})-(\ref{3f}) the result
\beqn
R_{b/\tau}= 3\frac{m_b^2}{m_{\tau}^2} 
\frac{(1+\epsilon_b'-\epsilon_b/\tan\alpha)^2}
{(1+\epsilon_b'+\epsilon_b\tan\beta)^2} 
\frac{(1+\epsilon_{\tau}'+\epsilon_{\tau}\tan\beta)^2}
{(1+\epsilon_{\tau}'-\epsilon_{\tau}/\tan\alpha)^2}
(1+\omega)
\label{4b}
\eeqn
where $\epsilon_{b,\tau}$ and $\epsilon_{b,\tau}'$ are defined so that 
\beqn
\epsilon_{b,\tau}= \Delta h_{b,\tau}/h_{b,\tau} ~~
\epsilon_{b,\tau}'= \delta h_{b,\tau}/h_{b,\tau} 
\eeqn
On neglecting the $\tau$ correction and $\epsilon_b'$ correction
this result agrees with Eq.(11) of the first paper of Ref.\cite{babu1998}.
 Similarly in the limit of vanishing phases one finds for $R_{b/c}$ the
result
\beqn
R_{b/c}= \frac{m_b^2}{m_{c}^2} (\tan\alpha \tan\beta)^2  
\frac{(1+\epsilon_b'-\epsilon_b \cot\alpha)^2}
{(1+\epsilon_b'+\epsilon_b\tan\beta)^2} 
\frac{(1+\epsilon_{c}'+\epsilon_{c}\cot\beta)^2}
{(1+\epsilon_{c}'-\epsilon_{c}\tan\alpha)^2}
\label{4c}
\eeqn
Again neglecting the $\epsilon_b'$ and $\epsilon_{c}'$ terms 
the above result agrees with Eq.(19) of the first paper of Ref.\cite{babu1998}.
We note, however, that in general the $\epsilon'$ term is not 
necessarily negligible.
 
\section{Numerical Analysis}
In this section we discuss the size of the supersymmetric 
corrections including CP phases to  $\Delta R_{b/\tau}$
and $\Delta R_{b/c}$ discussed in Sec.3.
We carry out the analysis in MSSM. Since the general
parameter space of MSSM is rather large we shall limit
our selves to a more constrained set for the purpose
of this numerical study.
We shall use for our parameter space the set
$m_A$, $m_0$, $m_{1/2}$, $|A_0|$, $\tan\beta$, $\theta_{\mu}$,
$\alpha_{A_0}$, $\xi_1$, $\xi_2$ and $\xi_3$. The parameter
$\mu$ is determined via radiative breaking of the electroweak
symmetry. The other sparticle masses are obtained from this set
using the renormalization group equations evolving the GUT parameters
from the GUT scale down to the electroweak scale.
 We then use the sparticle spectrum
generated by the above method to compute the supersymmetric  
corrections to the Higgs decays. 
 As mentioned above the phases chosen here for the numerical analysis
are $\xi_1$, $\xi_2$, $\xi_3$, $\alpha_{A0}$ and $\theta_{\mu}$.
This choice provides a convenient set of phases because of their 
appearance in the expressions for  $R_{ij}$, $\Delta h_f$ and $\delta h_f$.
By examining these expressions one finds
linear combinations of the phases listed above that enter the analysis.
Such linear combinations have been classified 
for EDMs in Ref.\cite{inmssm}.
Thus not all the phases are independent and absorbing one of these 
phases by redefinition of the other phases in these linear
combinations will not change the numerical analysis, i.e, the effect 
of CP phases on Higgs decay. However, for the numerical
analysis here it is more convenient to work with the set listed  above.

\begin{figure}
\hspace*{-0.6in}
\centering
\includegraphics[width=12cm,height=16cm]{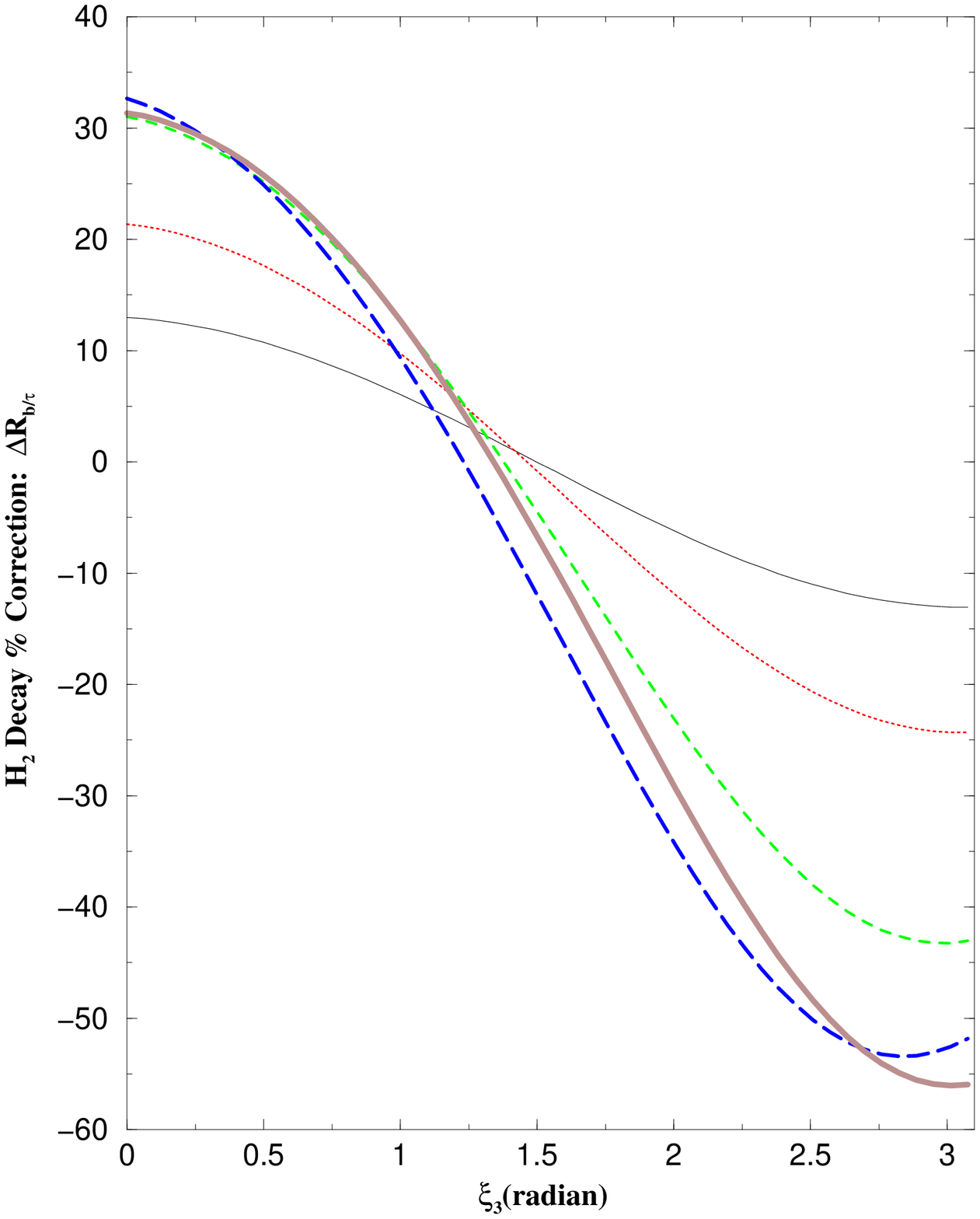}
\caption{ Plot of $\Delta R_{b/\tau}$ for the decay of
$H_2$ Higgs  as a function of the phase $\xi_3$.
 The other input parameters are:  
$m_A=200$ GeV, 
$m_0=m_{\frac{1}{2}}=200$ GeV,  $\xi_1=.5$, $\xi_2=.7$, 
$\theta_{\mu}=.1$, $\alpha_{A_0}=1.0$, and $|A_0|=4$.
The curves in descending order at the point
$\xi_3=\pi$ correspond to 
$\tan\beta =5, 10, 20, 30, 50$.  All angles here 
and in succeeding figure captions are in radians.}
\label{Ddelrbtau}
\end{figure}

\begin{figure}
\hspace*{-0.6in}
\centering
\includegraphics[width=12cm,height=16cm]{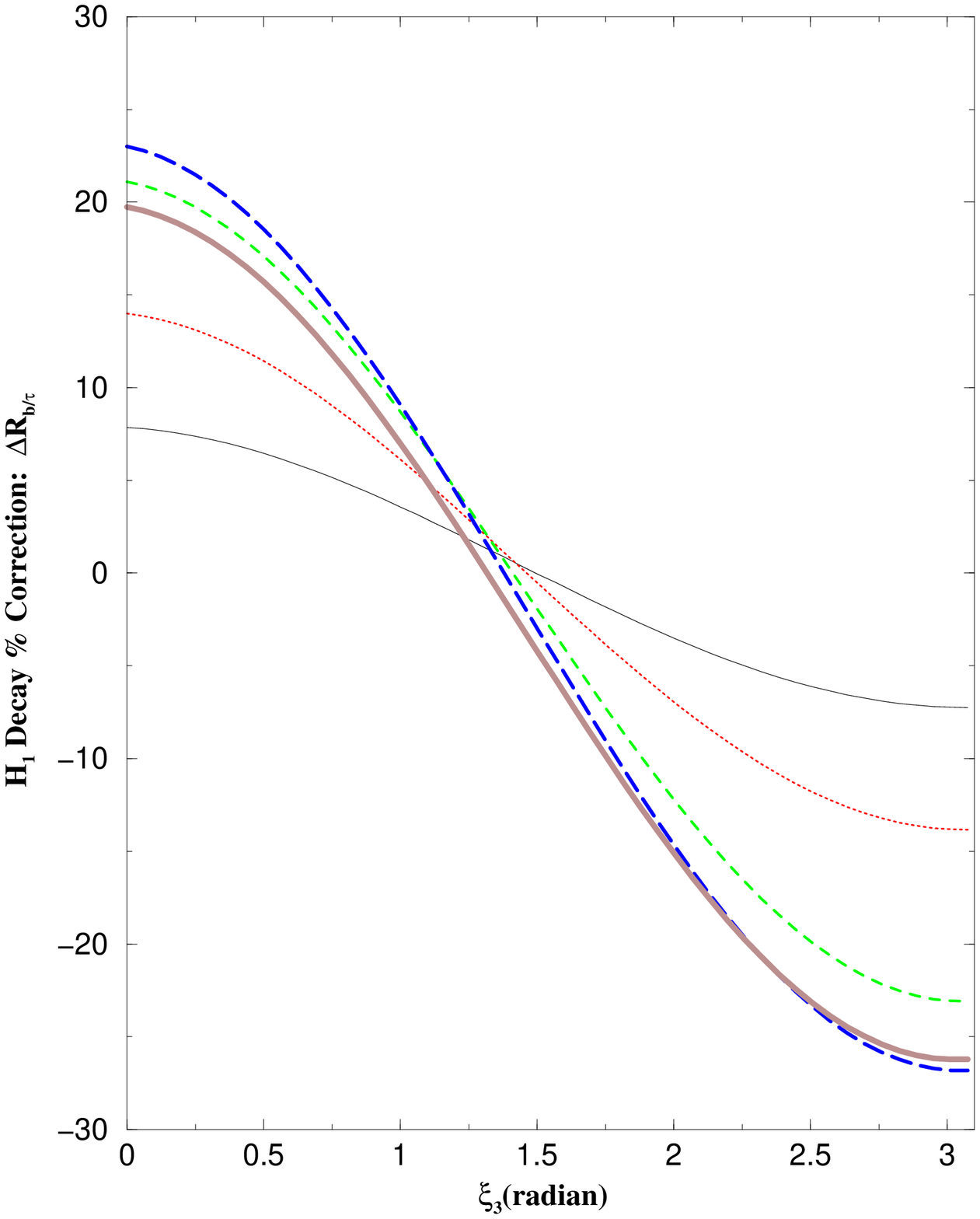}
\caption{ Plot of $\Delta R_{b/\tau}$ for the decay of
$H_1$ Higgs  as a function of the phase $\xi_3$.
 The other input parameters are:  
$m_A=200$ GeV,
$m_0=m_{\frac{1}{2}}=200$ GeV,  $\xi_1=.5$, $\xi_2=.7$, 
$\theta_{\mu}=.1$, $\alpha_{A_0}=1.0$, and $|A_0|=4$.
The curves in descending order at the point
$\xi_3= \pi$ correspond to 
$\tan\beta =5, 10, 20, 50, 30$.}
\label{Ddelrbtau1}
\end{figure}

\begin{figure}
\hspace*{-0.6in}
\centering
\includegraphics[width=12cm,height=16cm]{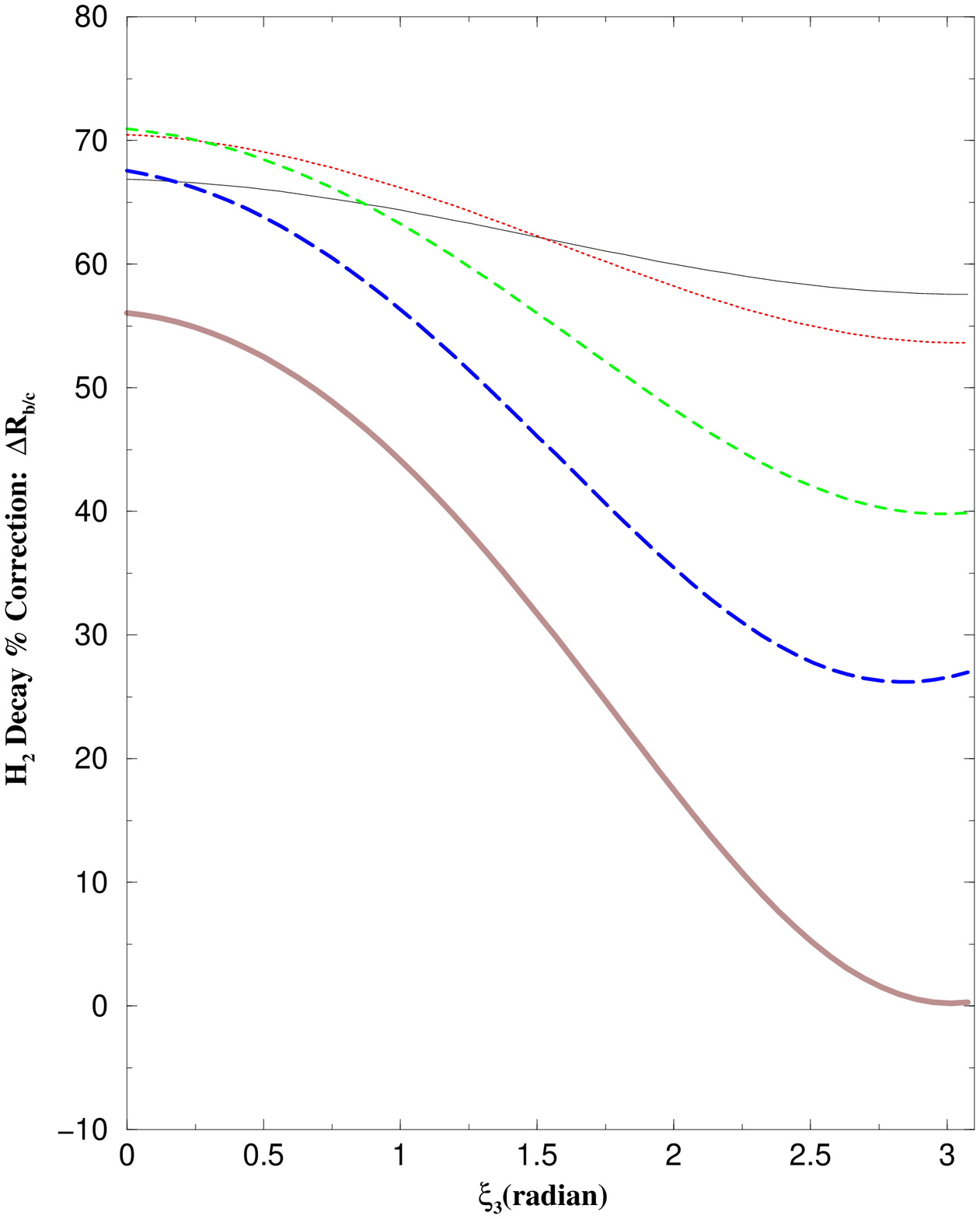}
\caption{ Plot of $\Delta R_{b/c}$ for the decay of
$H_2$ Higgs  as a function of the phase $\xi_3$.
 The other input parameters are:  
$m_A=200$ GeV,
$m_0=m_{\frac{1}{2}}=200$ GeV,  $\xi_1=.5$, $\xi_2=.7$, 
$\theta_{\mu}=.1$, $\alpha_{A_0}=1.0$, and $|A_0|=4$.
The curves in descending order at the point
$\xi_3=\pi$ correspond to 
$\tan\beta =5, 10, 20, 30, 50$.}
\label{Ddelrbc}
\end{figure}

\begin{figure}
\hspace*{-0.6in}
\centering
\includegraphics[width=12cm,height=16cm]{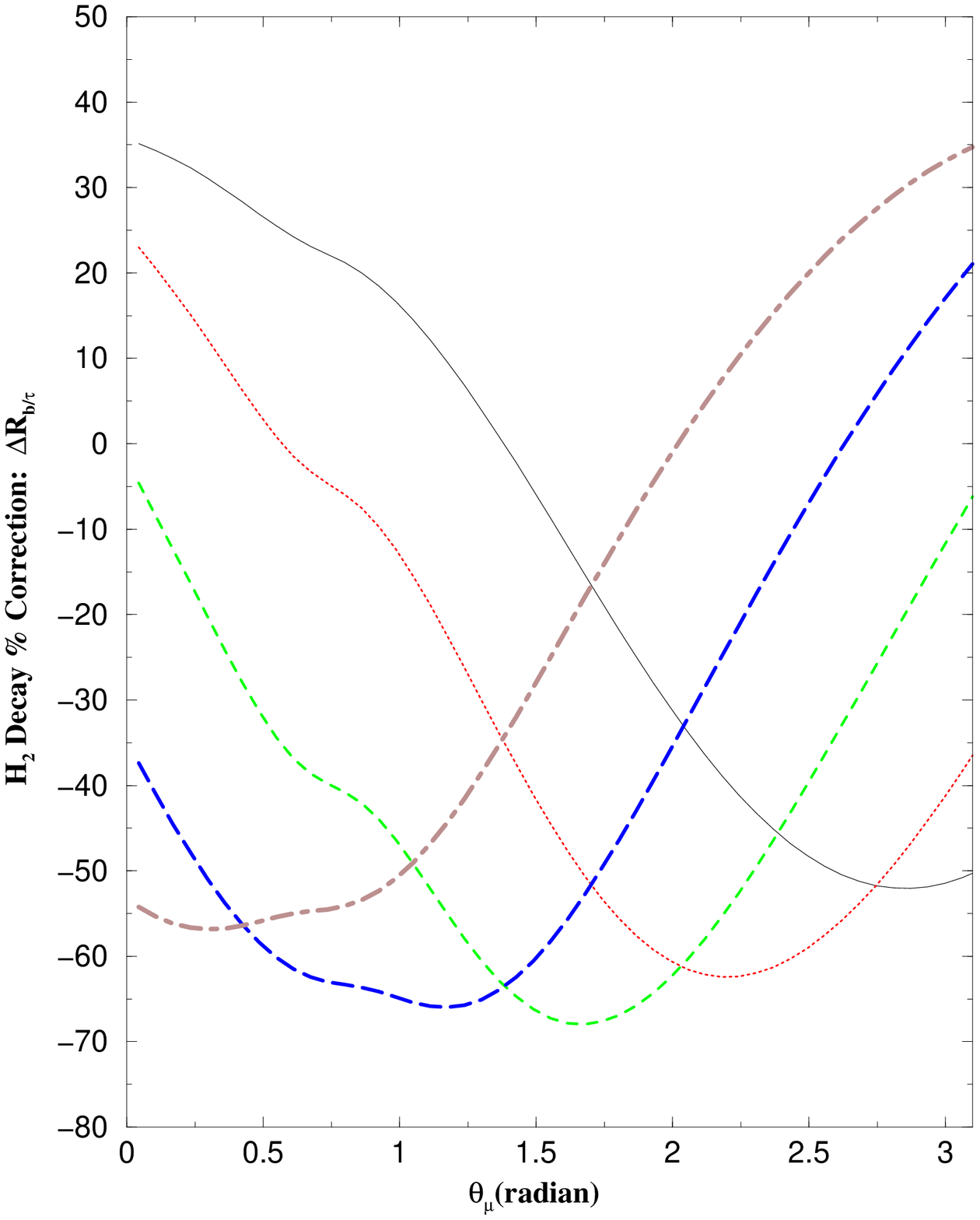}
\caption{ Plot of $\Delta R_{b/\tau}$ for the decay of
$H_2$ Higgs  as a function of the phase $\theta_{\mu}$.
 The other input parameters are:
$m_A=200$ GeV,
$m_0=m_{\frac{1}{2}}=300$,  
 $\xi_1=.5$, $\xi_2=.7$, $\alpha_{A_0}=1.0$, $|A_0|=5$,
and $\tan\beta =20$.
The curves in descending  order at $\theta_{\mu}=0$ correspond to 
  $\xi_3=0,.75,1.5,2.25,3$}
\label{Gdelrbtau}
\end{figure}

\begin{figure}
\hspace*{-0.6in}
\centering
\includegraphics[width=12cm,height=16cm]{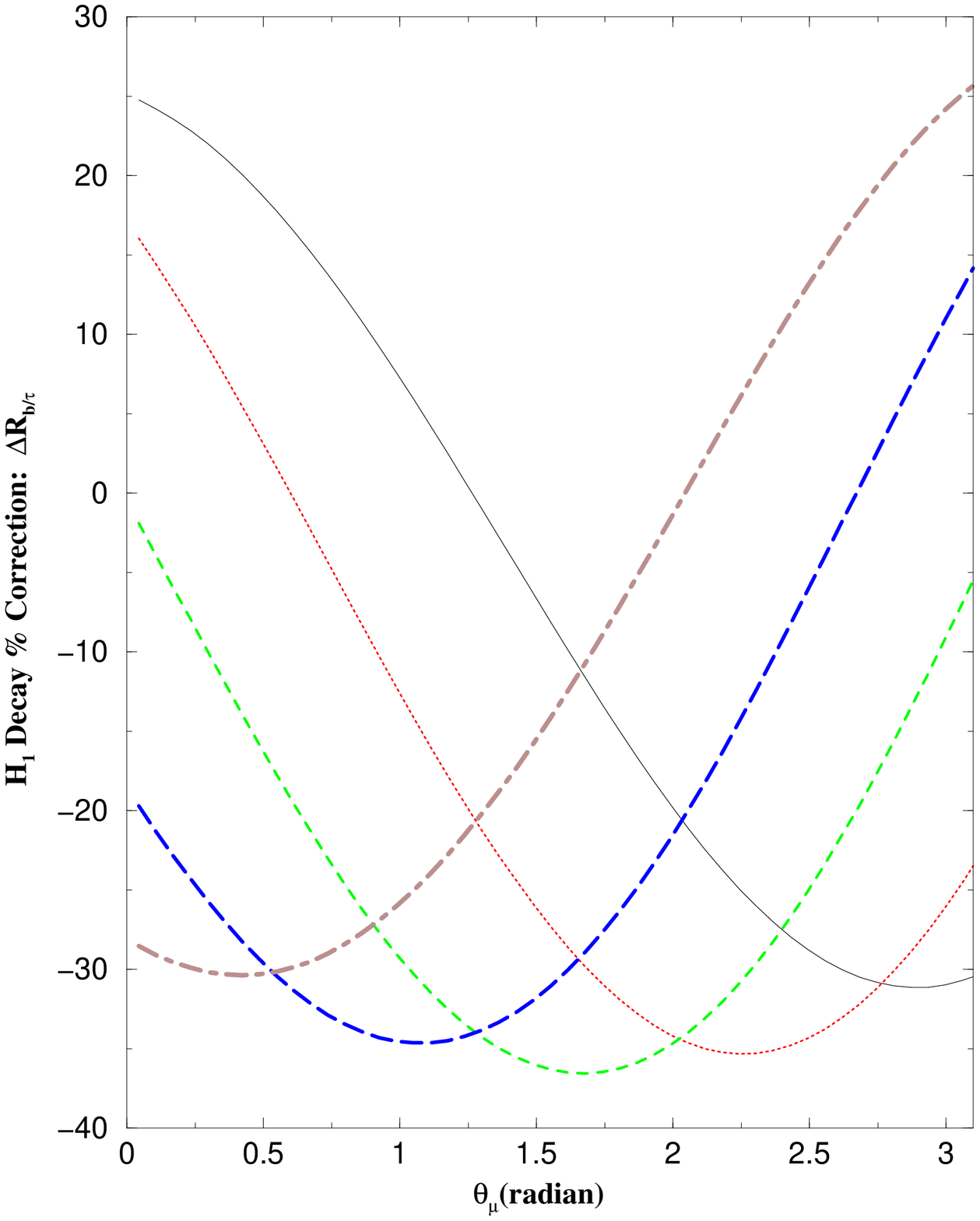}
\caption{ Plot of $\Delta R_{b/\tau}$ for the decay of
$H_1$ Higgs  as a function of the phase $\theta_{\mu}$.
The other input parameters are:
$m_A=200$ GeV,
$m_0=m_{\frac{1}{2}}=300$,  
 $\xi_1=.5$, $\xi_2=.7$, $\alpha_{A_0}=1.0$, $|A_0|=5$,
and $\tan\beta =20$.
The curves in descending  order  at $\theta_{\mu}=0$  correspond to 
  $\xi_3=0,.75,1.5,2.25,3$} 
\label{Gdelrbtau1}
\end{figure}

\begin{figure}
\hspace*{-0.6in}
\centering
\includegraphics[width=12cm,height=16cm]{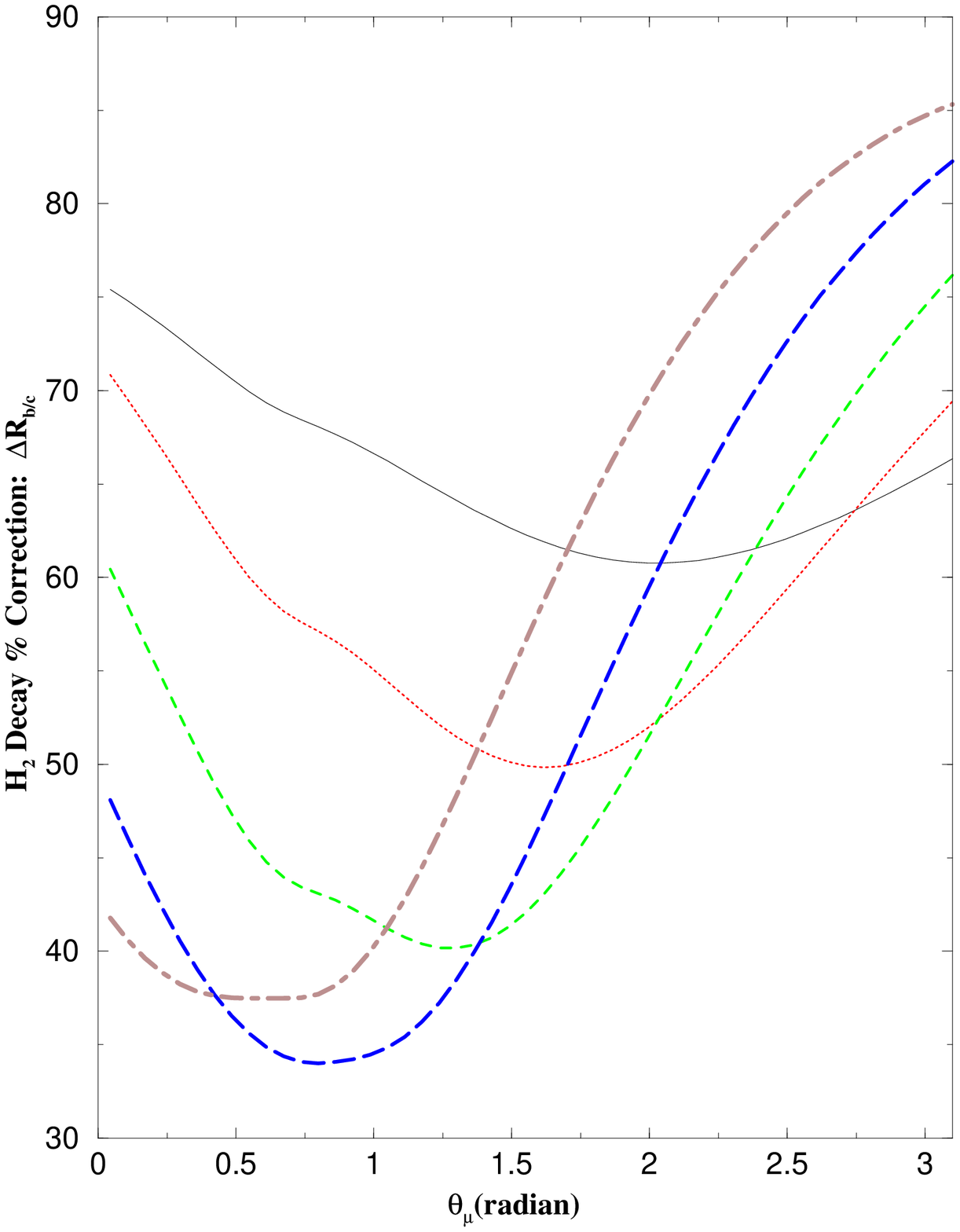}
\caption{ Plot of $\Delta R_{b/c}$ for the decay of
$H_2$ Higgs  as a function of the phase $\theta_{\mu}$.
The other input parameters are:
$m_A=200$ GeV,
$m_0=m_{\frac{1}{2}}=300$,  
 $\xi_1=.5$, $\xi_2=.7$, $\alpha_{A_0}=1.0$, $|A_0|=5$,
and $\tan\beta =20$.
The curves in descending  order  at $\theta_{\mu}=0$  correspond to 
  $\xi_3=0,.75,1.5,2.25,3$} 
\label{Gdelrbc}
\end{figure}

In Fig.~\ref{Ddelrbtau} we 
give a plot of $\Delta R_{b/\tau}$ for the decay of 
Higgs $H_2$ as given by Eq.(21) as a function of the phase
$\xi_3$ for  values of $\tan\beta$ ranging from 5-50. 
The analysis shows a very sharp dependence of $\Delta R_{b/\tau}$ 
on $\xi_3$. Thus $\xi_3$ affects not only the magnitude
of $\Delta R_{b/\tau}$ but also its sign depending on the
value of $\xi_3$.  A similar analysis holds for $\Delta R_{b/\tau}^{H_1}$
 where $ R_{b/\tau}^{H_1}$ is given by Eq.(54).
 Fig.~\ref{Ddelrbtau1} gives a plot of  
 $\Delta R_{b/\tau}^{H_1}$ as a function of $\xi_3$ for all the same inputs
as for Fig.~\ref{Ddelrbtau}. In this case also one finds a very
sharp dependence of $\Delta R_{b/\tau}^{H_1}$ on $\xi_3$. 
The analysis of $\Delta R_{b/\tau}^{H_3}$ follows a similar pattern
as $\Delta R_{b/\tau}^{H_1}$ and is not exhibited. 
As discussed in Sec.4 the quantity $\Delta R_{b/c}$ is also of 
 interest where $ R_{b/c}$ is given by Eq.(30). 
In Fig.~\ref{Ddelrbc} we give a plot of $\Delta R_{b/c}$ 
as a function of $\xi_3$ for all the same inputs as in 
the analysis of Fig.~\ref{Ddelrbtau}. Here also we find a very 
substantial dependence of $\Delta R_{b/c}$ on $\xi_3$. 
Next we analyse the variations of $\Delta R_{b/\tau}$ and 
$\Delta R_{b/c}$ as a function of $\theta_{\mu}$. 
In Fig.~\ref{Gdelrbtau} a plot of $\Delta R_{b/\tau}$ is given 
as function of $\theta_{\mu}$ for various values of $\xi_3$.
Similar plots for $\Delta R_{b/\tau}^{H_1}$ and for $\Delta R_{b/c}$
are given in Fig.~\ref{Gdelrbtau1} and Fig.~\ref{Gdelrbc}. 
Together 
Figs. \ref{Gdelrbtau}, ~\ref{Gdelrbtau1} and Fig.~\ref{Gdelrbc}
show  a rapid dependence of $\Delta R_{b/\tau}$,
 $\Delta R_{b/\tau}^{H_1}$ and of $\Delta R_{b/c}$ on $\theta_{\mu}$ 
 in addition to their sharp dependence on $\xi_3$.

\begin{figure}
\hspace*{-0.6in}
\centering
\includegraphics[width=12cm,height=16cm]{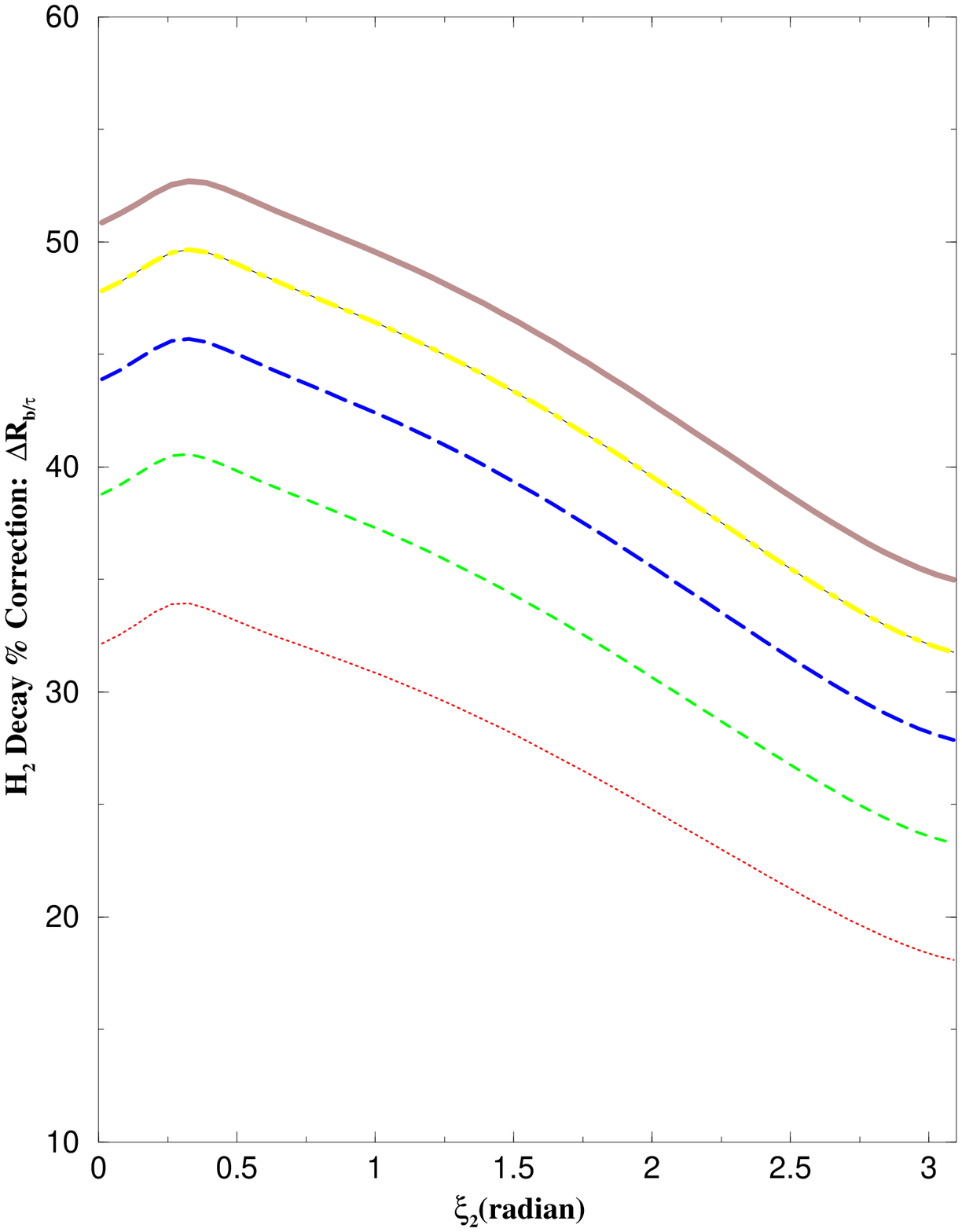}
\caption{ Plot of $\Delta R_{b/\tau}$ for the decay of
$H_2$ Higgs  as a function of the phase $\xi_2$.
 The other input parameters are:
$m_A=200$ GeV,  
 $\xi_1=.5$, $\xi_3=0$, 
$\theta_{\mu}=.1$, $\alpha_{A_0}=1.0$, $|A_0|=5$,
and $\tan\beta =50$.
The curves in ascending order correspond to 
$m_0=m_{\frac{1}{2}}=200,250,300,350,400$ GeV.}
\label{Fdelrbtau}
\end{figure}

\begin{figure}
\hspace*{-0.6in}
\centering
\includegraphics[width=12cm,height=16cm]{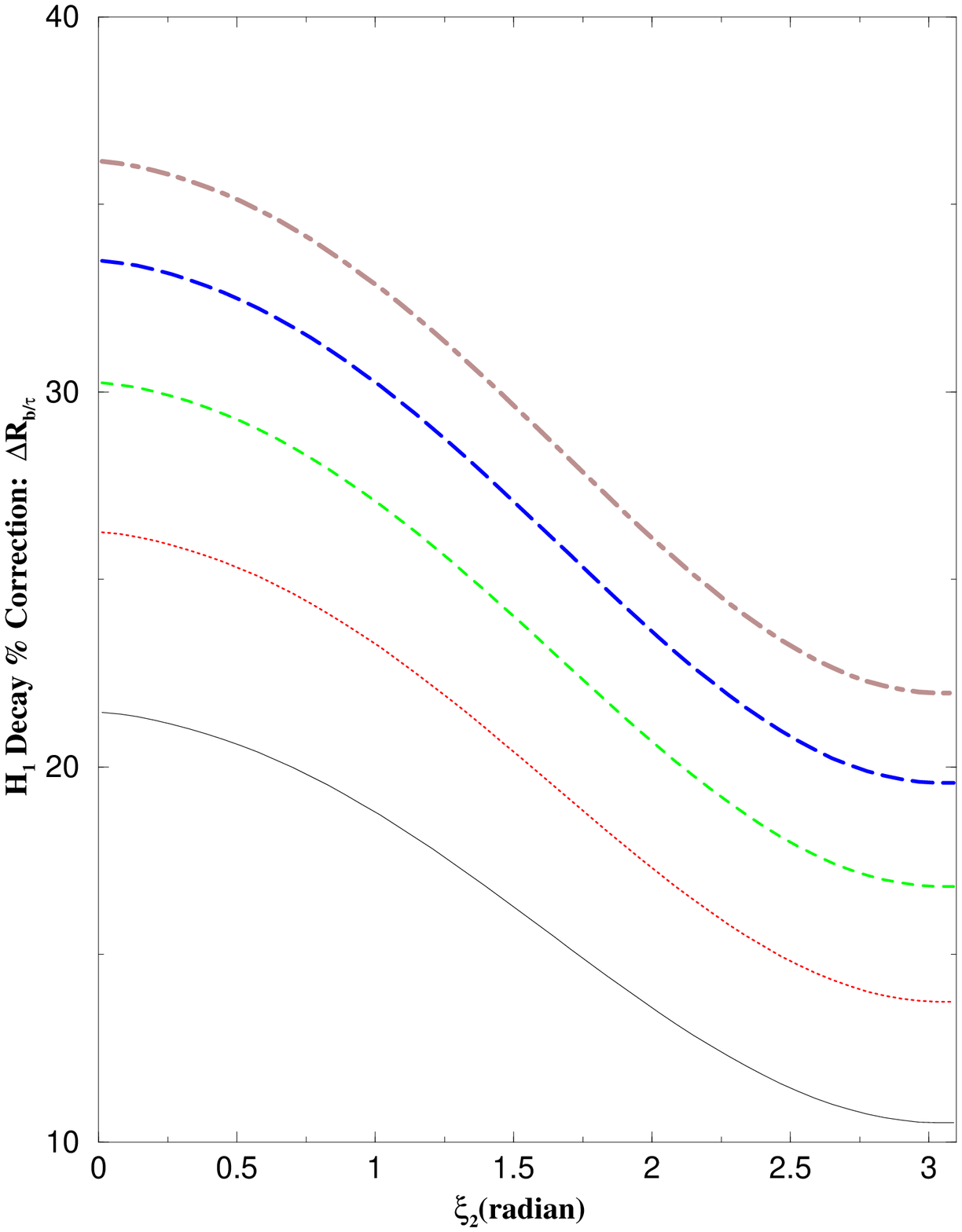}
\caption{ Plot of $\Delta R_{b/\tau}$ for the decay of
$H_1$ Higgs  as a function of the phase $\xi_2$.
 The other input parameters are:
$m_A=200$ GeV,  
 $\xi_1=.5$, $\xi_3=0$, 
$\theta_{\mu}=.1$, $\alpha_{A_0}=1.0$, $|A_0|=5$,
and $\tan\beta =50$.
The curves in ascending order correspond to 
$m_0=m_{\frac{1}{2}}=200,250,300,350,400$ GeV.}
\label{Fdelrbtau1}
\end{figure}

\begin{figure}
\hspace*{-0.6in}
\centering
\includegraphics[width=12cm,height=16cm]{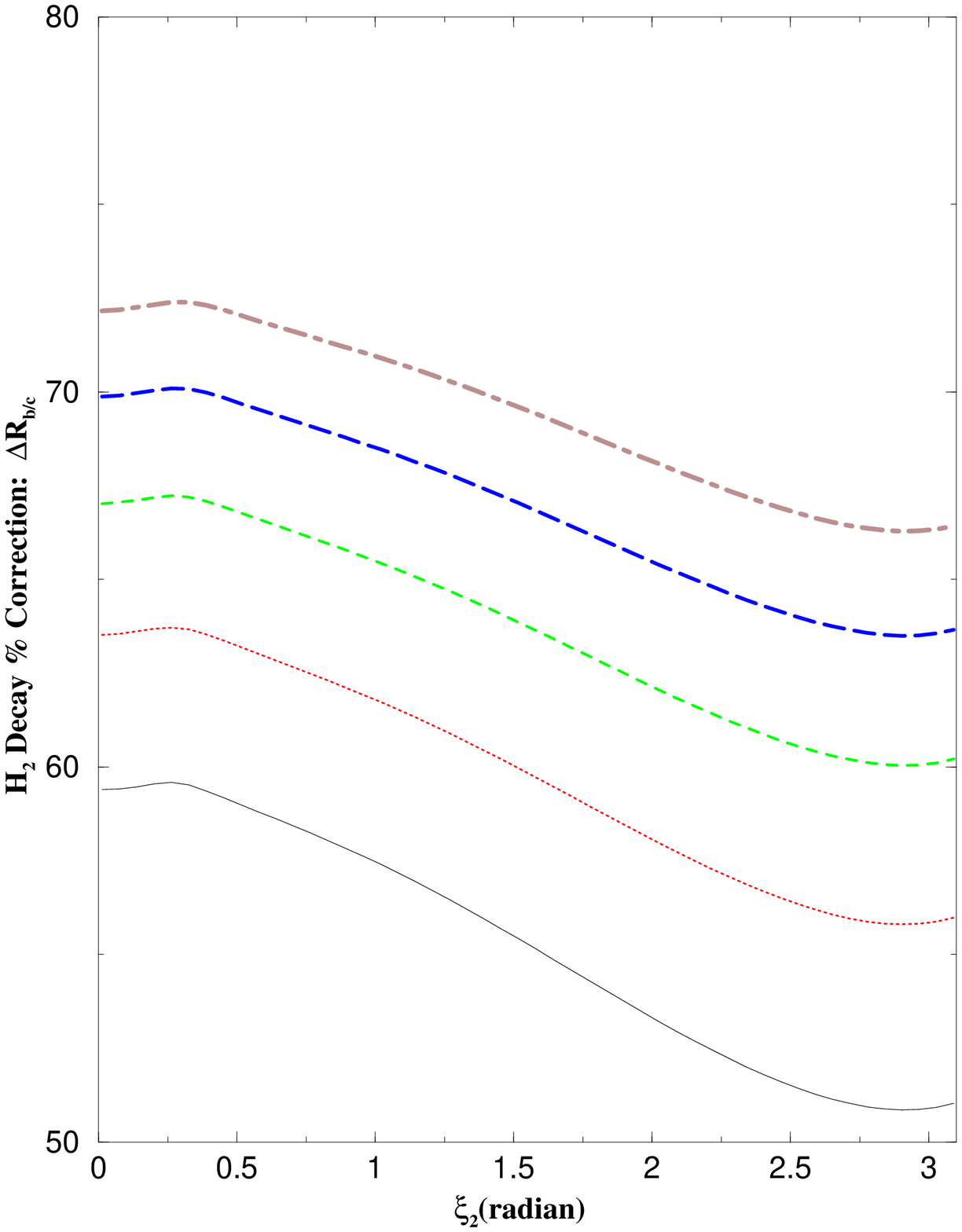}
\caption{ Plot of $\Delta R_{b/c}$ for the decay of
$H_2$ Higgs  as a function of the phase $\xi_2$.
 The other input parameters are:
$m_A=200$ GeV,  
 $\xi_1=.5$, $\xi_3=0$, 
$\theta_{\mu}=.1$, $\alpha_{A_0}=1.0$, $|A_0|=5$,
and $\tan\beta =50$.
The curves in ascending order correspond to 
$m_0=m_{\frac{1}{2}}=200,250,300,350,400$ GeV.}
\label{Fdelrbc}
\end{figure}

The analysis given in Fig.~\ref{Ddelrbtau}, Fig.~\ref{Ddelrbtau1}
and Fig.~\ref{Ddelrbc}, and also in 
Figs. \ref{Gdelrbtau}, ~\ref{Gdelrbtau1} and Fig.~\ref{Gdelrbc}
 include CP effects arising from 
both supersymmetric QCD and supersymmetric electroweak effects.
Next we study the dependence of the branching ratios on the
electroweak CP phase $\xi_2$. In  Fig.~\ref{Fdelrbtau} we give 
an analysis of $\Delta R_{b/\tau}$ for the decay of Higgs $H_2$
as a function of $\xi_2$ for values of $m_0,m_{\frac{1}{2}}$ 
ranging from 200 GeV -400 GeV. Here we see that the dependence
of $\Delta R_{b/\tau}$ on $\xi_2$ is not as strong as it is on
$\xi_3$. Nonetheless the $\xi_2$ dependence of  $\Delta R_{b/\tau}$  
 is still quite substantial as the variations in $\Delta R_{b/\tau}$
can be as much as 40\%. In  Fig.~\ref{Fdelrbtau1} we exhibit the
dependence of $\Delta R_{b/\tau}^{H_1}$  on $\xi_2$ 
and one finds that once again the dependence  though not
as strong as the $\xi_3$  dependence is still quite substantial
 and one can get variations of as much as 40\% over the 
 full range of $\xi_2$. 
The analysis of $\Delta R_{b/\tau}^{H_3}$ is very similar and
is not displayed. In Fig.~\ref{Fdelrbc} the dependence of 
$\Delta R_{b/c}$ on $\xi_2$ for the  $H_2$ decay is given. 
Again although the dependence of $\Delta R_{b/c}$ on $\xi_2$ 
is not as strong as on $\xi_3$ it is still quite substantial 
as one finds that $\Delta R_{b/c}$ can vary about 20\% over the
full range of $\xi_2$.

\begin{figure}
\hspace*{-0.6in}
\centering
\includegraphics[width=12cm,height=16cm]{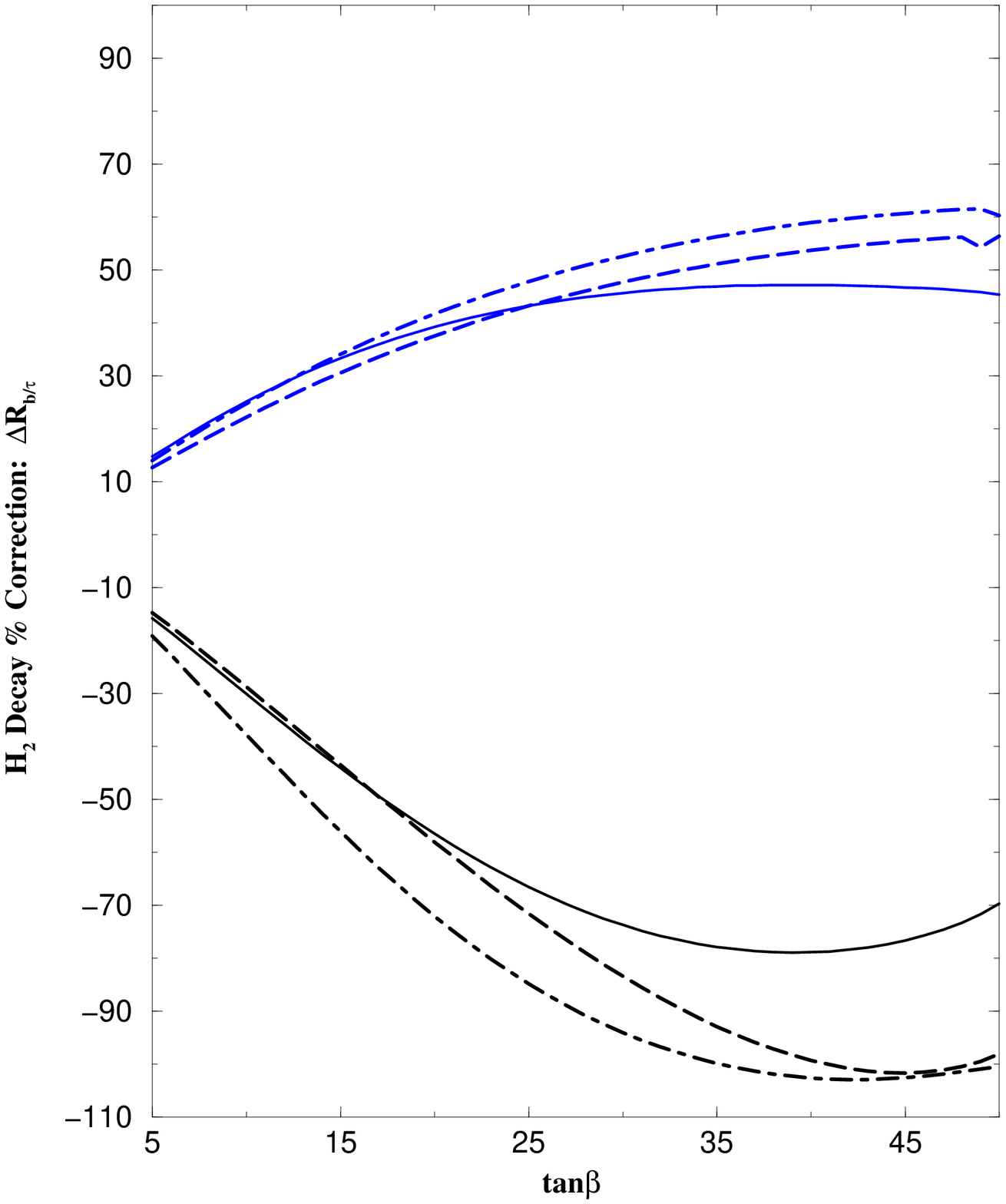}
\caption{ Plot  $\Delta R_{b/\tau}$ for the decay of $H_2$ Higgs
 as a function of $\tan\beta$. The  three lower curves are for the 
cases (i), (ii) and (iii) as follows: 
(i)$m_A=200$ GeV,  $m_0=m_{\frac{1}{2}}=300$ GeV, $A_0=4$, $\alpha_{A_0}=1$,
  $\xi_1=.5$, $\xi_2=.659$, $\xi_3=.633$, $\theta_{\mu}=2.5$ (solid) ; (ii) 
$m_A=200$ GeV,  
$m_0=m_{\frac{1}{2}}=555$ GeV, $A_0=4$, $\alpha_{A_0}=2$,
  $\xi_1=.6$, $\xi_2=.653$, $\xi_3=.672$,$\theta_{\mu}=2.5$ (dot-dashed); 
  (iii) $m_A=200$ GeV,
    $m_0=m_{\frac{1}{2}}=480$ GeV, $A_0=3$, $\alpha_{A_0}=.8$,
  $\xi_1=.4$, $\xi_2=.668$, $\xi_3=.6$, $\theta_{\mu}=2.5$ (long-dashed).
    The edm constraints including the $H_g^{199}$ are 
  satisfied for the above curves at $\tan\beta =50$ as shown in Table 1. 
  The three similar curves in
  the upper half plane are for the three cases above when the phases
  are all set to zero.}
\label{Adelrbtau}
\end{figure}

\begin{figure}
\hspace*{-0.6in}
\centering
\includegraphics[width=12cm,height=16cm]{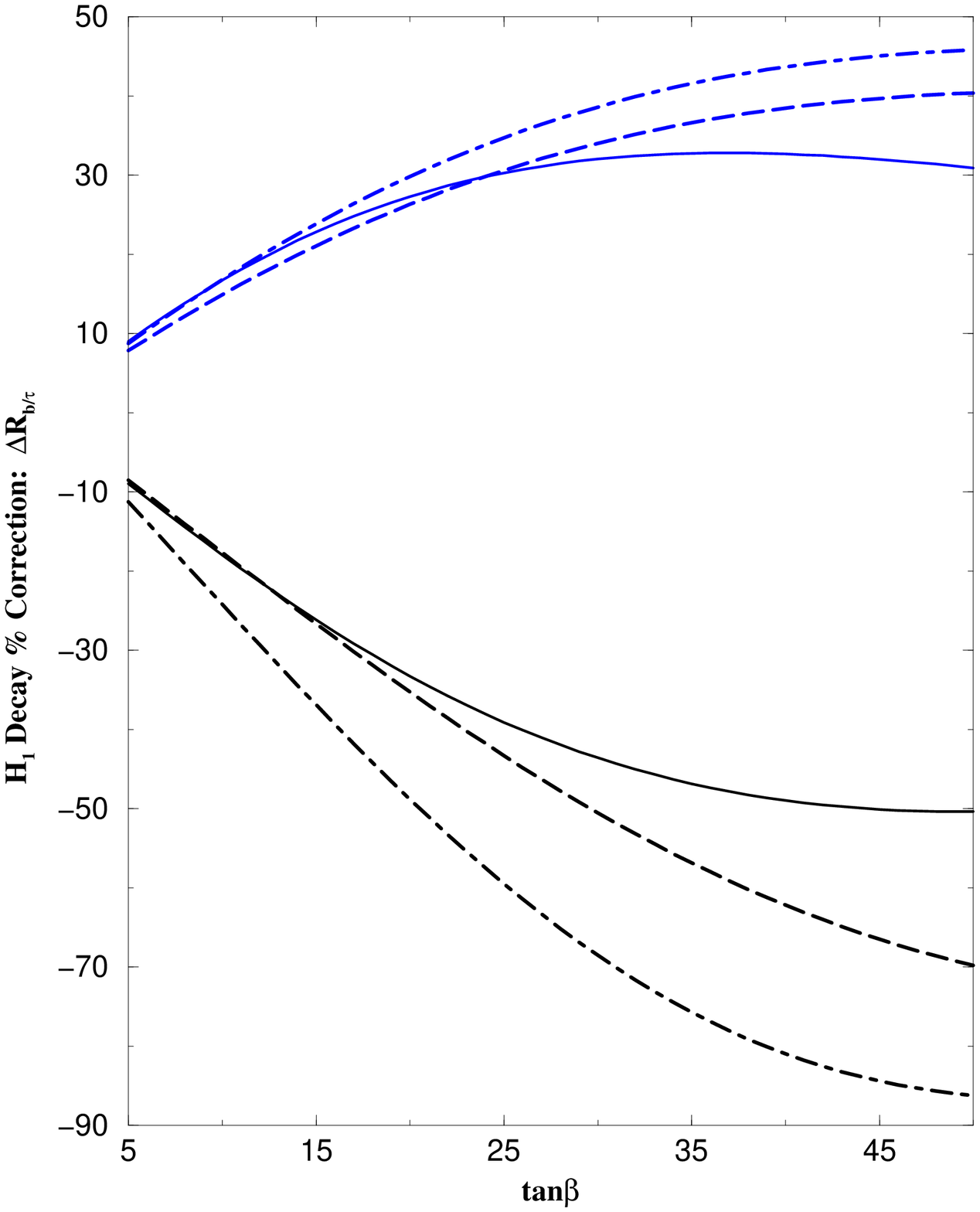}
\caption{ Plot  $\Delta R_{b/\tau}$ for the decay of $H_1$ Higgs
 as a function of $\tan\beta$.
 The  three lower curves are for the 
cases (i), (ii) and (iii) as follows: 
(i)$m_A=200$ GeV,  
$m_0=m_{\frac{1}{2}}=300$ GeV, $A_0=4$, $\alpha_{A_0}=1$,
  $\xi_1=.5$, $\xi_2=.659$, $\xi_3=.633$, $\theta_{\mu}=2.5$ (solid); (ii) 
$m_A=200$ GeV,  
$m_0=m_{\frac{1}{2}}=555$ GeV, $A_0=4$, $\alpha_{A_0}=2$,
  $\xi_1=.6$, $\xi_2=.653$, $\xi_3=.672$,$\theta_{\mu}=2.5$ (dot-dashed);
   (iii) $m_A=200$ GeV,
    $m_0=m_{\frac{1}{2}}=480$ GeV, $A_0=3$, $\alpha_{A_0}=.8$,
  $\xi_1=.4$, $\xi_2=.668$, $\xi_3=.6$, $\theta_{\mu}=2.5$ (long-dashed).
    The edm constraints including the $H_g^{199}$ are 
  satisfied for the above curves at $\tan\beta =50$ as shown in Table 1. 
  The three similar curves in
  the upper half plane are for the three cases above when the phases
  are all set to zero.}
\label{Adelrbtau1}
\end{figure}

\begin{figure}
\hspace*{-0.6in}
\centering
\includegraphics[width=12cm,height=16cm]{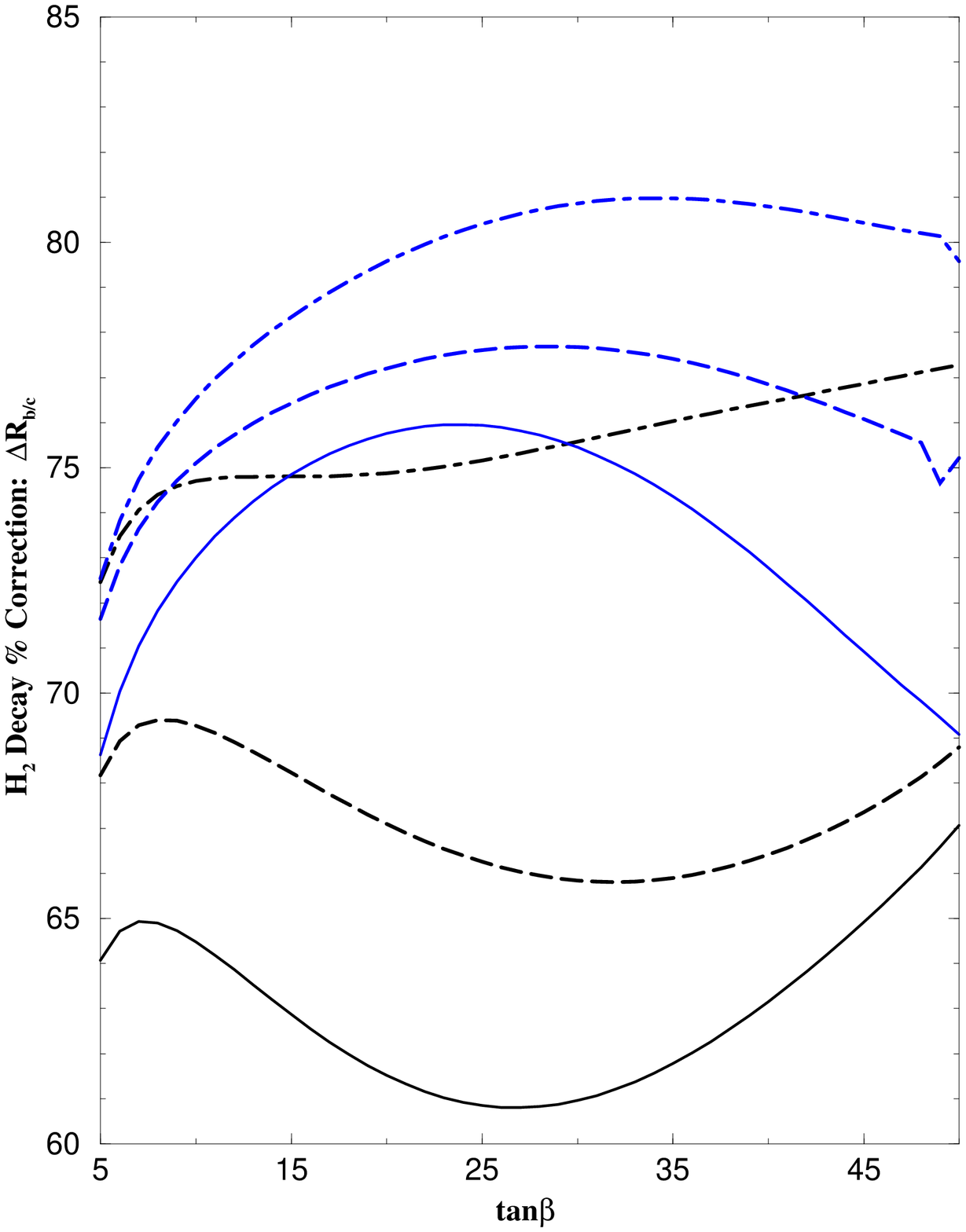}
\caption{ Plot  $\Delta R_{b/c}$ for the decay of $H_2$ Higgs
 as a function of $\tan\beta$.
 The  three lower curves are for the 
cases (i), (ii) and (iii) as follows: 
(i)$m_A=200$ GeV,  
$m_0=m_{\frac{1}{2}}=300$ GeV, $A_0=4$, $\alpha_{A_0}=1$,
  $\xi_1=.5$, $\xi_2=.659$, $\xi_3=.633$, $\theta_{\mu}=2.5$ (solid); (ii) 
$m_A=200$ GeV,  
$m_0=m_{\frac{1}{2}}=555$ GeV, $A_0=4$, $\alpha_{A_0}=2$,
  $\xi_1=.6$, $\xi_2=.653$, $\xi_3=.672$,$\theta_{\mu}=2.5$ (dot-dashed); (iii) 
$m_A=200$ GeV,    
$m_0=m_{\frac{1}{2}}=480$ GeV, $A_0=3$, $\alpha_{A_0}=.8$,
  $\xi_1=.4$, $\xi_2=.668$, $\xi_3=.6$, $\theta_{\mu}=2.5$ (long-dashed).
    The edm constraints including the $H_g^{199}$ are 
  satisfied for the above curves at $\tan\beta =50$ as shown in Table 1.
  The three similar curves in
  the upper half plane are for the three cases above when the phases
  are all set to zero.}
\label{Adelrbc}
\end{figure}

In Fig.~\ref{Adelrbtau} we give a plot of $\Delta R_{b/\tau}$ as a 
function of $\tan\beta$ for three different inputs which for the
largest value of $\tan\beta$ satisfy the edm constraints  including
the $H_g^{199}$ edm constraint.
(The $H_g^{199}$ edm constraint translates into 
$C_{Hg}=|d^C_d -d^C_u -0.012 d^C_s| < 3.0 \times 10^{-26} cm$\cite{olive,inhg199}.)
 The lower set of curves are 
for the cases  when phases are included while similar upper curves
are without phases. First one finds the effect of phases is  so large
as to not only affect the magnitude of $\Delta R_{b/\tau}$ but also
affect its sign. Further, one finds that  $\Delta R_{b/\tau}$ is 
strongly dependent on $\tan\beta$ for cases with and 
without phases. An identical analysis for $\Delta R_{b/\tau}^{H_1}$
is given in Fig.~\ref{Adelrbtau1} where $\Delta R_{b/\tau}^{H_1}$
is plotted as a function of $\tan\beta$ for exactly the same 
set of inputs as in Fig.~\ref{Adelrbtau}. One finds once again 
a very similar behavior in that both the magnitude and the sign of
$\Delta R_{b/\tau}^{H_1}$ are affected and also one finds  a very 
strong dependence on $\tan\beta$. The analysis of 
$\Delta R_{b/\tau}^{H_3}$ is very similar to that of 
$\Delta R_{b/\tau}^{H_1}$ and is not given. In Fig.~\ref{Adelrbc}
we give a plot of  $\Delta R_{b/c}$ as a function of $\tan\beta$ 
again for the same identical inputs as in Fig.~\ref{Adelrbtau}.
In this case also the effects of phases though smaller than in the
case of $\Delta R_{b/\tau}$ are still substantial in that the 
CP phase effects can be up to 30\%. Further, one finds about 
20\% variation due to variations in $\tan\beta$.

\begin{table}[h]
\begin{center}
\caption{EDM constraints for the points of $\tan\beta=50$ of Figs. 11
, 12 and 13.}
\begin{tabular}{|l|l|l|l|}
\hline
Case & $|d_e| e.cm$ &  $|d_n| e.cm$ &  $C_{Hg} cm$  \\
\hline
(i)   & $1.67 \times 10^{-27}$ & $1.59 \times 10^{-27}$ & $1.18 \times 10^{-27}$  \\
\hline
(ii)  & $6.05 \times 10^{-28}$ & $3.47 \times 10^{-27}$ & $1.29 \times 10^{-26}$  \\
\hline
(iii) & $2.14 \times 10^{-27}$
 & $8.90 \times 10^{-28}$ & $1.25 \times 10^{-26}$\\  
\hline
\end{tabular}
\end{center}
\end{table}

\section{Conclusions}
In this paper we have computed the effects of CP phases on the
Higgs boson decays. In supersymmetry after spontaneous breaking 
one is left with three neutral Higgs bosons which in the absence of
CP phases consist of two CP even Higgs and one CP odd Higgs.
With the inclusion of CP phases the Higgs mass eigen states
are no longer CP eigen states but rather admixtures of CP 
even and CP odd states when loop corrections to the Higgs
boson masses are included. Further, inclusion of loop corrections
to the quarks and lepton masses are in general dependent on CP 
phases and these effects can be very significant for the case
of the b quark mass. Additionally inclusion of CP dependent loop
effects on the quark and lepton masses modify the vertices
involving the quarks and leptons and the Higgs mass eigen states
and these modifications also affect the Higgs boson decays. 
 In this paper we have computed the effects of
CP phases on the decays of the  light and the heavy Higgs boson decays
to $b\bar b$ and to $\tau\bar  \tau$. Specifically we computed the
deviation of $R_{b/\tau}$ from the standard model prediction in
the presence of CP phases. We find that these effects can
be rather large. Thus $\Delta R_{b/\tau}$  can vary by as much as 
100\% or more due to the variation in the $\theta_{\mu}$ and
$\xi_3$ while variations with $\xi_2$ are  relatively small although
still substantial. A similar analysis was also carried out for 
$\Delta R_{b/c}$ and one finds that the variations of $\Delta R_{b/c}$
 with phases are also substantial.
The analysis presented here points to the possible detection
of CP phases via accurate measurement of the branching ratios
provided we have sufficient constraints on the remaining SUSY
input parameters from other experiment. Thus the decays of the
Higgs if measured with sufficient accuracy may 
provide a signal for the presence of both supersymmetry and CP phases.\\
\noindent
{\bf Acknowledgments}\\ 
This research was also supported in part by NSF grant PHY-0139967\\
\noindent
{\bf Appendix A: Analysis of {\boldmath {$\Delta h_c$ and  $\delta h_c$.}}} \\
Following the same technique as given in Ref..\cite{in2003} we give below 
explicit expressions for $\Delta h_c$ and $\delta h_c$.
Thus the  c quark mass at the Z scale is given by 
\beq 
m_c(M_Z)=h_c(M_Z)\frac{v}{\sqrt 2}\sin\beta(1+\Delta_c) 
\eeq
where $\Delta_c$ gives the loop correction to the c quark mass $m_c$. 
and $\Delta_c$ is given by  
\beqn
\Delta_c = (Re\frac{\Delta h_c}{h_c} cot\beta +
Re\frac{\delta h_c}{h_c})
\eeqn
An analysis of $\Delta h_c$ to  one loop order gives
\beqn
\Delta h_c = - \sum_{\it i =1}^2 \sum_{j=1}^2 \frac{2\alpha_s}{3\pi}
 e^{-i\xi_3}m_{\tilde g} 
F_{cij}^* D_{c1i}^{*} D_{c2j} 
f(m_{\tilde g}^2,m_{\tilde c_{\it i}}^2,m_{\tilde c_j}^2)\nonumber\\
-\sum_{i=1}^2\sum_{j=1}^2\sum_{k=1}^2    
g^2 H_{cij}^*\{U_{k1}^*D_{s1i}^* -k_s U_{k2}^* D_{s2i}^*\}
(k_c V_{k2}^*D_{s1j})
\frac{m_{\chi_k^+}}{16\pi^2}
f(m_{\chi_k^+}^2,m_{\tilde s_i}^2,m_{\tilde s_j}^2)\nonumber\\
-\sum_{i=1}^2\sum_{j=1}^2\sum_{k=1}^2    
g^2 K_{ji}^*\{U_{i1}^*D_{s1k}^* -k_s U_{i2}^* D_{s2k}^*\}
(k_c V_{j2}^*D_{s1k})
\frac{m_{\chi_i^+}m_{\chi_j^+}}{16\pi^2}
f(m_{\tilde s_k}^2,m_{\chi_i^+}^2,m_{\chi_j^+}^2)\nonumber\\
-\sum_{i=1}^2\sum_{j=1}^2 \sum_{k=1}^4 
2F_{cij}^* 
\{\alpha_{ck}D_{c1j}-\gamma_{ck}D_{c2j}\} 
   \{\beta_{ck}^*D_{c1i}^*+\alpha_{ck}D_{c2i}^*\} 
  \frac{m_{\chi_k^0}}{16\pi^2}
f(m_{\chi_k^0}^2,m_{\tilde c_i}^2,m_{\tilde c_j}^2)\nonumber\\
-\sum_{i=1}^4\sum_{j=1}^4 \sum_{k=1}^2 
2\Delta_{ij}^* 
\{\alpha_{cj}D_{c1k}-\gamma_{cj}D_{c2k}\} 
   \{\beta_{ci}^*D_{c1k}^*+\alpha_{ci}D_{c2k}^*\} 
  \frac{m_{\chi_i^0} m_{\chi_j^0}}{16\pi^2}
f(m_{\tilde c_k}^2, m_{\chi_i^0}^2, m_{\chi_j^0}^2 ) 
\eeqn
where 
\beqn
\alpha_{ck} =\frac{g_2m_cX_{4k}}{2m_W\sin\beta}\nonumber\\
\beta_{ck}=eQ_cX_{1k}^{'*} +\frac{g}{\cos\theta_W} X_{2k}^{'*}
(T_{3c}-Q_c\sin^2\theta_W)\nonumber\\
\gamma_{ck}=eQ_c X_{1k}'-\frac{gQ_c\sin^2\theta_W}{\cos\theta_W}
X_{2k}'
\eeqn
\beqn
k_{c(s)}=\frac{m_{c(s)}}{\sqrt 2 m_W \sin\beta (\cos\beta)}
\eeqn
and where $Q_c=\frac{2}{3}$ and $T_{3c}=\frac{1}{2}$. 
In the above $D_{cij}$ is the matrix that diagonalizes the c squark 
$mass^2$
matrix  and $\tilde c_i$ are the c squark 
mass eigen states so that
\beqn
\tilde c_L=\sum_{i=1}^{2} D_{c1i} \tilde c_i,~~~~~ 
\tilde c_R=\sum_{i=1}^{2} D_{c2i} \tilde c_i
\eeqn
Similarly  $D_{sij}$ is the matrix that diagonalizes the s squark 
$mass^2$
matrix  and $\tilde s_i$ are the s squark 
mass eigen states so that
\beqn
\tilde s_L=\sum_{i=1}^{2} D_{s1i} \tilde s_i,~~~~~ 
\tilde s_R=\sum_{i=1}^{2} D_{s2i} \tilde s_i
\eeqn
Further, U and V are the matrices that diagonalize the chargino 
mass matrix and the matrix X diagonalizes the neutralino mass
matrix and the elements of the $X'$ matrix are  defined by
\beqn
X'_{1k}=X_{1k}\cos\theta_W +X_{2k}\sin\theta_W,~~~
X'_{2k}=-X_{1k}\sin\theta_W +X_{2k}\cos\theta_W
\eeqn
where we are using a notation of Ref.\cite{in2003}.
$F_{cij}$ and $H_{cij}$ are defined by 
\beqn
\frac{F_{cij}}{\sqrt 2} =-\frac{gM_Z}{2\cos\theta_W} 
\{(\frac{1}{2} -\frac{2}{3}\sin^2\theta_W)
D^*_{c1i}D_{c1j} +\frac{2}{3} \sin^2\theta_W D^*_{c2i}D_{c2j}\}
\cos\beta\nonumber\\
+\frac{gm_c\mu}{2M_W\sin\beta} 
 D^*_{c1i}D_{c2j}
\eeqn
and 
\beqn
\frac{H_{cij}}{\sqrt 2} =-\frac{gM_Z}{2\cos\theta_W} 
\{(-\frac{1}{2} +\frac{1}{3}\sin^2\theta_W)
D^*_{s1i}D_{s1j} -\frac{1}{3} \sin^2\theta_W D^*_{s2i}D_{s2j}\}
\cos\beta\nonumber\\
-\frac{gm_s^2}{2M_W\cos\beta} 
[ D^*_{s1i}D_{s1j} + D^*_{s2i}D_{s2j}]
-\frac{gm_sm_0A_s}{2M_W\cos\beta} 
 D^*_{s2i}D_{s1j} 
\eeqn
while $K_{ij}$ and $\Delta_{ij}$ are defined by 
\beqn
\frac{K_{ij}}{\sqrt 2}= -\frac{g}{2} Q_{ji},~~~
\frac{\Delta_{ij}}{\sqrt 2}= -\frac{g}{2} Q_{ij}{''} 
\eeqn
where
\beqn
Q_{ij}= \sqrt{\frac{1}{2}} U_{i2}V_{j1}\nonumber\\
gQ^{''}_{ij}= \frac{1}{2} [ X_{3i}^* (gX_{2j}^* -g' X_{1j}^*) +
(i\leftarrow \rightarrow j) ]
\eeqn
Finally the function $f(m^2,m_i^2,m_j^2)$ is given by
\beqn
f(m^2,m_i^2,m_j^2)
= \frac {1}{(m^2-m_i^2) (m^2-m_j^2)(m_j^2-m_i^2)}
(m_j^2 m^2 ln\frac{m_j^2}{m^2} 
 +m^2 m_i^2ln\frac{m^2}{m_i^2} +m_i^2 m_j^2 ln\frac{m_i^2}{m_j^2})  
 \eeqn
 for the case $i\neq j$ and  
 \beqn
 f(m^2,m_i^2,m_j^2) =\frac {1}{(m_i^2-m^2)^2} (m^2 ln\frac{m_i^2}{m^2} 
 + (m^2-m_i^2))
\eeqn 
for the case i=j.
Similarly for $\delta h_c$ we find the result
\beqn
\delta h_c = - \sum_{\it i =1}^2 \sum_{j=1}^2 \frac{2\alpha_s}{3\pi} 
e^{-i\xi_3}m_{\tilde g} 
E_{cji} D_{c1i}^{*} D_{c2j} 
f(m_{\tilde g}^2,m_{\tilde c_{\it i}}^2,m_{\tilde c_j}^2)\nonumber\\
-\sum_{i=1}^2\sum_{j=1}^2\sum_{k=1}^2    
g^2 G_{cji}\{U_{k1}^*D_{s1i}^* -k_s U_{k2}^* D_{s2i}^*\}
(k_c V_{k2}^*D_{s1j})
\frac{m_{\chi_k^+}}{16\pi^2}
f(m_{\chi_k^+}^2,m_{\tilde s_i}^2,m_{\tilde s_j}^2)\nonumber\\
-\sum_{i=1}^2\sum_{j=1}^2\sum_{k=1}^2    
g^2 C_{ji}^*\{U_{i1}^*D_{s1k}^* -k_s U_{i2}^* D_{s2k}^*\}
(k_c V_{j2}^*D_{s1k})
\frac{m_{\chi_i^+}m_{\chi_j^+}}{16\pi^2}
f(m_{\tilde s_k}^2,m_{\chi_i^+}^2,m_{\chi_j^+}^2)\nonumber\\
-\sum_{i=1}^2\sum_{j=1}^2 \sum_{k=1}^4 
2E_{cji} 
\{\alpha_{ck}D_{c1j}-\gamma_{ck}D_{c2j}\} 
   \{\beta_{ck}^*D_{c1i}^*+\alpha_{ck}D_{c2i}^*\} 
  \frac{m_{\chi_k^0}}{16\pi^2}
f(m_{\chi_k^0}^2,m_{\tilde c_i}^2,m_{\tilde c_j}^2)\nonumber\\
-\sum_{i=1}^4\sum_{j=1}^4 \sum_{k=1}^2 
2\Gamma_{ij}^* 
\{\alpha_{cj}D_{c1k}-\gamma_{cj}D_{c2k}\} 
   \{\beta_{ci}^*D_{c1k}^*+\alpha_{ci}D_{c2k}^*\} 
  \frac{m_{\chi_i^0} m_{\chi_j^0}}{16\pi^2}
f(m_{\tilde c_k}^2, m_{\chi_i^0}^2, m_{\chi_j^0}^2 ) 
\eeqn
where $E_{cij}$ and $G_{cij}$ are given by 
\beqn
\frac{E_{cij}}{\sqrt 2} =\frac{gM_Z}{2\cos\theta_W} \{(\frac{1}{2} -\frac{2}{3}\sin^2\theta_W)
D^*_{c1i}D_{c1j} +\frac{2}{3} \sin^2\theta_W D^*_{c2i}D_{c2j}\}
\sin\beta\nonumber\\
-\frac{gm_c^2}{2M_W\sin\beta} 
[ D^*_{c1i}D_{c1j} + D^*_{c2i}D_{c2j}]
-\frac{gm_cm_0A_c}{2M_W\sin\beta} 
 D^*_{c2i}D_{c1j} 
\eeqn
and
\beqn
\frac{G_{cij}}{\sqrt 2} =\frac{gM_Z}{2\cos\theta_W} 
\{(-\frac{1}{2} +\frac{1}{3}\sin^2\theta_W)
D^*_{s1i}D_{s1j} -\frac{1}{3} \sin^2\theta_W D^*_{s2i}D_{s2j}\}
\sin\beta\nonumber\\
+\frac{gm_s\mu}{2M_W\cos\beta} D^*_{s1i}D_{s2j}
\eeqn
while $C_{ij}$ and $\Gamma_{ij}$ are defined by 
\beqn
\frac{C_{ij}}{\sqrt 2}= -\frac{g}{2\sin\beta} 
[ \frac{m_{\chi_i^+}}{2M_W} \delta_{ij}
-Q^*_{ij} \cos\beta - R^*_{ij} ] 
\eeqn
and 
\beqn
\frac{\Gamma_{ij}}{\sqrt 2}= 
-\frac{g}{2\sin\beta} [ \frac{m_{\chi_i^0}}{2M_W} \delta_{ij}
-Q^{''*}_{ij} \cos\beta
- R^{''*}_{ij}] 
\eeqn
where
\beqn
R_{ij}=\frac{1}{2M_W} [\tilde m_2^* U_{i1} V_{j1} + \mu^* U_{i2}
V_{j2}]\nonumber\\
 R_{ij}^{''}= \frac{1}{2M_W} [ \tilde m_1^* X_{1i}^*X_{1j}^* 
+ \tilde m_2^* X_{2i}^*X_{2j}^*
-\mu^* (X_{3i}^*X_{4j}^* + X_{4i}^*X_{3j}^*) ] 
\eeqn
~\\
\noindent
{\bf Appendix B: Decays of $H_1$ and $H_3$ Higgs bosons. }\\
We can repeat the same analysis for the heavy Higgs bosons $H_1$ and
$H_3$.
The effective interaction of the b quark with the Higgs mass
eigen state $H_1$ is given by 
\beqn
-{\cal {L}}_{int}^b= \bar b [C_b^{S'} + i\gamma_5 C_b^{P'}]b H_1\nonumber\\
C_b^{S'}= \tilde C_b^{S'} \cos\chi_b -\tilde C_b^{P'} \sin\chi_b\nonumber\\
 C_b^{P'}= \tilde C_b^{S'} \sin\chi_b +\tilde C_b^{P'} \cos\chi_b\nonumber\\
\sqrt 2 \tilde C_b^{S'} = Re (h_b +\delta h_b) R_{11} + [-Im (h_b + \delta h_b) 
 \sin\beta\nonumber\\  + Im (\Delta h_b)\cos\beta ] R_{13}
+ Re (\Delta h_b) R_{12}\nonumber\\
\sqrt 2 \tilde C_b^{P'} = - Im (h_b +\delta h_b) R_{11} + [-Re (h_b + \delta h_b)
\sin\beta\nonumber\\ + Re (\Delta h_b) \cos\beta ]R_{13}
- Im (\Delta h_b) R_{12} 
\label{B1}
\eeqn
The $\tau$ lepton has similar interactions. Thus 
\beqn
-{\cal {L}}_{int}^{\tau}= \bar \tau [C_{\tau}^{S'} + i\gamma_5 C_{\tau}^{P'}]\tau H_1 
\label{B2} 
\eeqn
and $C_{\tau}^{S'}$ and $C_{\tau}^{P'}$ can be obtained from Eq.~(\ref{B1})
by the interchange $b\rightarrow \tau$.  
 For the $H_1$ decays one finds
\beqn
R_{b/\tau}^{H_1} =\frac{BR(H_1\rightarrow \bar b b)}{BR(H_1\rightarrow \bar\tau\tau)}
\label{B3}
\eeqn
Using the interactions given by Eqs.~(\ref{B1})and ~(\ref{B2})
 we dertmine this to be
\beqn
R_{b/\tau}^{H_1} = 3\frac{m_b^2}{m_{\tau}^2} \frac{S_{\tau}}{S_b} 
\frac{T_{b}^{H_1}}{T_{\tau}^{H_1}} (1+\omega) 
\label{B4}
\eeqn
where
\beqn
T_f^{H_1}= (\frac{M_{H_1}^2-4m_f^2}{M_{H_1}^2})^{\frac{3}{2}}
(A_f^{S'})^2 +
(\frac{M_{H_1}^2-4m_f^2}{M_{H_1}^2})^{\frac{1}{2}}(A_f^{P'})^2
\label{B5}
\eeqn
where
\beqn
A_f^{S'}=\tilde A_f^{S'} \cos\chi_f - \tilde A_f^{P'} \sin\chi_f,~~
A_f^{P'}=\tilde A_f^{S'} \sin\chi_f +\tilde A_f^{P'} \cos\chi_f\nonumber\\
\tilde A_f^{S'}=[(1+Re\frac{\delta h_f}{h_f}) R_{11}  - 
Im\frac{\delta h_f}{h_f} R_{13} 
\sin\beta + Re\frac{\Delta h_f}{h_f}R_{12}
+ Im\frac{\Delta h_f}{h_f}R_{13}
\cos\beta]\nonumber\\
\tilde A_f^{P'}=
[-Im\frac{\delta h_f}{h_f} R_{11}  - (1+ Re\frac{\delta h_f}{h_f}) 
R_{13} \sin\beta
- Im\frac{\Delta h_f}{h_f}R_{12} + Re\frac{\Delta h_f}{h_f}R_{13}
\cos\beta]
\label{B6}
\eeqn

The interaction that governs the  $H_3$ decay is 
\beqn
-{\cal {L}}_{int}^b= \bar b [C_b^{S''} + i\gamma_5 C_b^{P''}]b H_3\nonumber\\
C_b^{S''}= \tilde C_b^{S''} \cos\chi_b -\tilde C_b^{P''} \sin\chi_b\nonumber\\
 C_b^{P''}= \tilde C_b^{S''} \sin\chi_b +\tilde C_b^{P''} \cos\chi_b\nonumber\\
\sqrt 2 \tilde C_b^{S''} = Re (h_b +\delta h_b) R_{31} + [-Im (h_b + \delta h_b) 
 \sin\beta\nonumber\\  + Im (\Delta h_b)\cos\beta ] R_{33}
+ Re (\Delta h_b) R_{32}\nonumber\\
\sqrt 2 \tilde C_b^{P''} = - Im (h_b +\delta h_b) R_{31} + [-Re (h_b + \delta h_b)
\sin\beta\nonumber\\ + Re (\Delta h_b) \cos\beta ]R_{33}
- Im (\Delta h_b) R_{32} 
\label{B8}
\eeqn
Similarly the $\tau$ lepton interaction with $H_3$ is given by 
\beqn
-{\cal {L}}_{int}^{\tau}= \bar \tau [C_{\tau}^{S''} + i\gamma_5 C_{\tau}^{P''}]\tau H_3 
\label{B9} 
\eeqn
and $C_{\tau}^{S''}$ and $C_{\tau}^{P''}$ can be obtained from Eq.~(\ref{B8})
by the interchange $b\rightarrow \tau$.  
 For the $H_3$ decays one finds
\beqn
R_{b/\tau}^{H_3} =\frac{BR(H_3\rightarrow \bar b b)}{BR(H_3\rightarrow \bar\tau\tau)}
\label{B10}
\eeqn
Using the interactions given in Eqs.~(\ref{B8}) and ~(\ref{B9})
 we dertmine this to be
\beqn
R_{b/\tau}^{H_3} = 3\frac{m_b^2}{m_{\tau}^2} \frac{S_{\tau}}{S_b} 
\frac{T_{b}^{H_3}}{T_{\tau}^{H_3}} (1+\omega) 
\label{B11}
\eeqn
where

\beqn
T_f^{H_3}= (\frac{M_{H_3}^2-4m_f^2}{M_{H_3}^2})^{\frac{3}{2}}
(A_f^{S''})^2 +
(\frac{M_{H_3}^2-4m_f^2}{M_{H_3}^2})^{\frac{1}{2}}(A_f^{P''})^2
\label{B12}
\eeqn
where
\beqn
A_f^{S''}=\tilde A_f^{S''} \cos\chi_f - \tilde A_f^{P''} \sin\chi_f,~~
A_f^{P''}=\tilde A_f^{S''} \sin\chi_f +\tilde A_f^{P''} \cos\chi_f\nonumber\\
\tilde A_f^{S''}=[(1+Re\frac{\delta h_f}{h_f}) R_{31}  - 
Im\frac{\delta h_f}{h_f} R_{33} 
\sin\beta + Re\frac{\Delta h_f}{h_f}R_{32}
+ Im\frac{\Delta h_f}{h_f}R_{33}
\cos\beta]\nonumber\\
\tilde A_f^{P''}=
[-Im\frac{\delta h_f}{h_f} R_{31}  - (1+ Re\frac{\delta h_f}{h_f}) 
R_{33} \sin\beta
- Im\frac{\Delta h_f}{h_f}R_{32} + Re\frac{\Delta h_f}{h_f}R_{33}
\cos\beta]
\label{B13}
\eeqn

%%%%%%%%%%%%%%%%%%%%%%%%%%%%%%%%

\newpage

\newpage

\newpage

\newpage

\newpage

\newpage

\newpage

\newpage

\newpage

\newpage

\newpage

\newpage

\end{document}